\documentclass[letterpaper,twocolumn,english,prc,aps,showpacs,superscriptaddress]{revtex4}

\usepackage{ae,aecompl}
\usepackage[T1]{fontenc}
\usepackage[latin9]{inputenc}
\usepackage{babel}
\usepackage{units}
\usepackage{amsmath}
\usepackage{amssymb}
\usepackage{graphicx}
\usepackage{subfigure}
\usepackage{tikz}
\usepackage{multirow}
\usepackage{ulem}

\makeatletter

\DeclareTextFontCommand{\helvetica}{\fontfamily{phv}\selectfont}

\makeatother

\begin{document}


\title{Understand the thermometry of hot nuclei\\
 from the energy spectra of light charged particles }

\author{E. Vient}
\thanks{vient@lpccaen.in2p3.fr; http://www.lpc-caen.in2p3.fr/}
\affiliation{Normandie Univ, ENSICAEN, UNICAEN, CNRS/IN2P3, LPC Caen, F-14000 Caen, France}

\author{L. Augey}
\affiliation{Normandie Univ, ENSICAEN, UNICAEN, CNRS/IN2P3, LPC Caen, F-14000 Caen, France}

\author{B. Borderie}
\affiliation{Institut de Physique Nucl\'eaire, CNRS/IN2P3, Univ. Paris-Sud, Universit\'e Paris-Saclay, F-91406 Orsay cedex, France}

\author{A. Chbihi}
\affiliation{Grand Acc\'el\'erateur National d'Ions Lourds (GANIL), CEA/DRF-CNRS/IN2P3, Bvd. Henri Becquerel, 14076 Caen, France}

\author{D. Dell'Aquila}
\affiliation{Institut de Physique Nucl\'eaire, CNRS/IN2P3, Univ. Paris-Sud, Universit\'e Paris-Saclay, F-91406 Orsay cedex, France}
\affiliation{Dipartimento di Fisica 'E. Pancini' and Sezione INFN, Universit\'a di Napoli 'Federico II', I-80126 Napoli, Italy}

\author{Q. Fable}
\affiliation{Grand Acc\'el\'erateur National d'Ions Lourds (GANIL), CEA/DRF-CNRS/IN2P3, Bvd. Henri Becquerel, 14076 Caen, France}

\author{L. Francalanza}
\affiliation{Dipartimento di Fisica 'E. Pancini' and Sezione INFN, Universit\'a di Napoli 'Federico II', I-80126 Napoli, Italy}

\author{J.D. Frankland}
\affiliation{Grand Acc\'el\'erateur National d'Ions Lourds (GANIL), CEA/DRF-CNRS/IN2P3, Bvd. Henri Becquerel, 14076 Caen, France}

\author{E. Galichet}
\affiliation{Institut de Physique Nucl\'eaire, CNRS/IN2P3, Univ. Paris-Sud, Universit\'e Paris-Saclay, F-91406 Orsay cedex, France}
\affiliation{Conservatoire National des Arts et M\'etiers, F-75141 Paris Cedex 03, France}

\author{D. Gruyer}
\affiliation{Dipartimento di Fisica, Universit\'a di Firenze, via G. Sansone 1, I-50019 Sesto Fiorentino (FI), Italy}
\affiliation{Normandie Univ, ENSICAEN, UNICAEN, CNRS/IN2P3, LPC Caen, F-14000 Caen, France}

\author{D. Guinet}
\affiliation{IPNL/IN2P3 et Universit\'e de Lyon/Universit\'e Claude Bernard Lyon 1, 43 Bd du 11 novembre 1918 F69622 Villeurbanne Cedex, France}

\author{M. Henri}
\affiliation{Normandie Univ, ENSICAEN, UNICAEN, CNRS/IN2P3, LPC Caen, F-14000 Caen, France}

\author{M. La Commara}
\affiliation{Dipartimento di Fisica 'E. Pancini' and Sezione INFN, Universit\'a di Napoli 'Federico II', I-80126 Napoli, Italy}

\author{E. Legou\'ee}
\affiliation{Normandie Univ, ENSICAEN, UNICAEN, CNRS/IN2P3, LPC Caen, 14000 Caen, France}

\author{G. Lehaut}
\affiliation{Normandie Univ, ENSICAEN, UNICAEN, CNRS/IN2P3, LPC Caen, 14000 Caen, France}

\author{N. Le Neindre}
\affiliation{Normandie Univ, ENSICAEN, UNICAEN, CNRS/IN2P3, LPC Caen, F-14000 Caen, France}

\author{I. Lombardo}
\affiliation{Dipartimento di Fisica 'E. Pancini' and Sezione INFN, Universit\'a di Napoli 'Federico II', I-80126 Napoli, Italy}
\affiliation{INFN - Sezione Catania, via Santa Sofia 64, 95123 Catania, Italy}

\author{O. Lopez}
\affiliation{Normandie Univ, ENSICAEN, UNICAEN, CNRS/IN2P3, LPC Caen, F-14000 Caen, France}

\author{L. Manduci}
\affiliation{Ecole des Applications Militaires de l'Energie Atomique, BP 19 50115, Cherbourg Arm\'ees, France}
\affiliation{Normandie Univ, ENSICAEN, UNICAEN, CNRS/IN2P3, LPC Caen, F-14000 Caen, France}

\author{P. Marini}
\affiliation{CEA, DAM, DIF, F-91297 Arpajon, France}

\author{M. P\^arlog}
\affiliation{Normandie Univ, ENSICAEN, UNICAEN, CNRS/IN2P3, LPC Caen, F-14000 Caen, France}
\affiliation{Hulubei National Institute for R$\And$D in Physics and Nuclear Engineering (IFIN-HH), P.O.BOX MG-6, RO-76900 Bucharest-M\`agurele, Romania}

\author{M. F. Rivet}
\thanks{deceased}
\affiliation{Institut de Physique Nucl\'eaire, CNRS/IN2P3, Univ. Paris-Sud, Universit\'e Paris-Saclay, F-91406 Orsay cedex, France}

\author{E. Rosato}
\thanks{deceased}
\affiliation{Dipartimento di Fisica 'E. Pancini' and Sezione INFN, Universit\'a di Napoli 'Federico II', I-80126 Napoli, Italy}

\author{R. Roy}
\affiliation{Laboratoire de Physique Nucl\'eaire, Universit\'e Laval, Qu\'ebec, Canada G1K 7P4}

\author{P. St-Onge}
\affiliation{Laboratoire de Physique Nucl\'eaire, Universit\'e Laval, Qu\'ebec, Canada G1K 7P4}
\affiliation{Grand Acc\'el\'erateur National d'ions Lourds (GANIL), CEA/DRF-CNRS/IN2P3, Bvd. Henri Becquerel, 14076 Caen, France}

\author{G. Spadaccini}
\affiliation{Dipartimento di Fisica 'E. Pancini' and Sezione INFN, Universit\'a di Napoli 'Federico II', I-80126 Napoli, Italy}

\author{G. Verde}
\affiliation{Institut de Physique Nucl\'eaire, CNRS/IN2P3, Univ. Paris-Sud, Universit\'e Paris-Saclay, F-91406 Orsay cedex, France}
\affiliation{INFN - Sezione Catania, via Santa Sofia 64, 95123 Catania, Italy}

\author{M. Vigilante}
\affiliation{Dipartimento di Fisica 'E. Pancini' and Sezione INFN, Universit\'a di Napoli 'Federico II', I-80126 Napoli, Italy}

\vspace{0.5cm}
\collaboration{INDRA collaboration}
\noaffiliation

\date{\today}
 
\begin{abstract}
In the domain of Fermi energy, the hot nucleus temperature can be determined by using the energy spectra of evaporated light charged particles. But this method of measurement is not without difficulties both theoretical and experimental. The presented study aims to disentangle the respective influences of different factors on the quality of this measurement :  the physics,  the detection (a 4$\pi$ detector array as INDRA) and the experimental procedure. This analysis demonstrates the possibility of determining from an energy spectrum, with an accuracy of about 10 \%, the true apparent temperature felt by a given type of particle emitted by a hot nucleus. Three conditions are however necessary : have a perfect detector of particles, an important statistics and very few secondary emissions. According to the GEMINI event generator, for hot nuclei of intermediate mass, only deuterons and tritons could fill these conditions. This temperature can allow to trace back to the initial temperature by using an appropriate method. This determination may be better than 15 \%.  With a real experimental device, an insufficient angular resolution and topological distortions caused by the detection can damage spectra to the point to make very difficult a correct determination of the apparent temperature. The experimental reconstruction of the frame of the hot nucleus may also be responsible for this deterioration. 
\end{abstract}

\pacs{24.10.-i ; 24.10.Pa ; 24.60.-k ; 25.70.-z ; 25.70.Gh}

\keywords{Heavy ions; Hot nuclear matter; Thermometry; Excitation energy; Caloric curves;
4 $\pi$ detection array; Methodology; Experimental errors;}
\keywords{Suggested keywords}
\maketitle

\section{\label{sec1}Introduction}
The concept of temperature was introduced in nuclear physics by V.Weisskopf in a statistical model describing the neutron emission by a hot nucleus \cite{weisskopf1}. In theoretical nuclear physics, two types of temperature \cite{ericson1,morrissey1} are often defined: a temperature called "nuclear", which would correspond to a "microcanonical" temperature \cite{gulminelli1} and a temperature called "thermodynamic" \cite{ericson1,morrissey1}, which would correspond to a canonical or grand canonical temperature. These two temperatures become identical only at the thermodynamic limit. For a hot nucleus, an isolated thermodynamic system with a small number of nucleons, this last condition seems never be fulfilled \cite{gross1}. The  canonical or grand canonical ensemble must be therefore used cautiously depending of the observable to study or of the event sorting chosen \cite{gross2, borderie1}. 
 
Many methods, used to measure the temperature of a hot nucleus, are nevertheless based on the hypothesis that the thermodynamics of hot nuclei can be described either by a canonical or by a grand canonical ensemble \cite{kelic1, borderie1}. Then, we can assert that the measurement of the "nuclear" temperature will be more significant if the studied hot nucleus is constituted of a large number of nucleons.

The concept of temperature needs a priori to admit that nuclei have had enough time to thermalize during the collisions. From the theoretical side it was shown that energy relaxation can be partially or totaly fulfilled in the Fermi energy domain, depending on incident energy \cite{toepffer1, cugnon1}. Experimental results have confirmed these expectations \cite{jouan1}. Moreover we can also mention that the chaotic character of collisions involved favors a large covering of statistical partitions when an homogeneous event sample is studied  \cite{borderie2}.

Even though the conceptual and theoretical bases of the temperature in nuclear physics can seem fragile, its use and importance in nuclear thermodynamics  \cite{bonche1, levit1, besprosvany1,pochodzalla1, natowitz1} brought us to want to make a "metrology" of a classic nuclear thermometry. We chose to concentrate on the thermometry done by using energetic spectra of light charged particles. The measured temperature by this way seems strongly correlated to the "nuclear " temperature of the hot nucleus \cite{lefevre1}.  A similar study has already been made for different methods of temperature measurement, but it did not at all take into account the influence of the experimental device on these measurements  as done here \cite{siwek1}.

In the first section, we will remind briefly these theoretical bases and we will explain how this thermometry is actually applied. We will define their intrinsic limitations, dependent on physics and assumptions on which it is based. In the second section,  we will estimate the effective quality of the temperature measurements by determining relative uncertainties existing on these measures. We will try to identify the different causes of these uncertainties by making a specific methodological analysis. 
The main tools for this study are the event generators GEMINI of R.Charity \cite{charity1, charity2, nicolis1} and SIMON, created by D.Durand \cite{durand1}.  These two codes are based on statistical models of evaporation.

To build spectra of evaporated light particles is a very difficult experimental task. We know that the reaction mechanisms have a fundamental importance when we want to characterize a hot nucleus. It is not obvious to disentangle what comes from the collision and what is really due to the de-excitation of the hot nucleus. 

The theory, on which is based the nuclear thermometry, assumes that the hot nucleus is isolated. But during the collisions of heavy ions, it is clear that this assumption is false. The process of evaporation will be necessarily disturbed by the Coulomb field of the other participants in the collision and perhaps even by their nuclear field, if they are very close \cite{gruyer1}. The use of the experimental data is far from being easy. We do not know the origin of the detected particles. There is a mixture between the particles coming indeed from the hot nucleus of interest and those coming from the pre-equilibrium or from the disintegration of another nucleus. 
 
It is also of course necessary to reconstruct the initial velocity of the emitting nucleus. This reconstruction of the initial frame is done by using a specific experimental procedure. It is important to understand its influence on the construction of the energy spectra and consequently on the measurement of the temperature. Moreover,  it is necessary to add the effect of the evaporation cascade. We will study thus what we will call the \textbf{method effect}.

The experimental device of detection has a certain efficiency. It presents a certain potential of identification of charges and masses, thresholds of detection and of identification, an angular resolution and a limited energy resolution, which will generate also distortions of the energy spectra. 
We will speak in this case about an experimental \textbf{filter effect}.

Another important aspect, often neglected, is the statistics. Indeed, the calorimetric studies require event and particle selections for different reasons. For a correct calorimetry, well detected events are needed. It is necessary also to try to eliminate in a proper way the contributions unevaporated by the studied nucleus.  In fact, these different selections can bring geometrical bias concerning the properties of the evaporated light particles \cite{Vient1}. All these different selections can generate important problems of statistics. The estimate of the temperature not being direct, it is necessary to use numerical methods of fit, whose quality depends on the statistics actually present in the experimental spectrum.

To try to understand the influence of each one of these factors, we will follow a specific protocol. 
 With an event generator, we will consider ourselves in the ideal physical case of an isolated hot nucleus whose initial excitation energy and initial velocity are known. This will allow us to get rid of the problems that poses the collision itself.
Initially, we will suppose that we have no detection effect. We will thus be able to apply our measurement method in "a perfect way" independently of the formation process of the hot nucleus and of the detection set-up. We will thus study the only effect of \textbf{the measurement method}.
 We then will try to determine the only influence of the experimental filter on the temperature measurement. For that, we will use the event generators and a software filter simulating the detection of the multi-detector INDRA \cite{copinet1}. The quality of this last one was validated for the reactions Xe + Sn at 50 A.MeV.
 
 For the evaporated particles actually detected, then, we will reconstruct the energy spectra in the real frame of the initial hot nucleus. Thus, we will judge  only \textbf{the filter effect}.
To complete, we will study the simulated and filtered events by reconstructing the initial frame by the usual experimental method \cite{Vient1,steck1}. 
Finally, we  will thus test the cumulative effect on the temperature measurement of the used method, of the experimental filter and of the reconstructing protocol of the hot nucleus frame, i.e. the total  \textbf{experimental effect}.
\section{Theoretical bases and practical application of the studied thermometry\label{sec2}}
\subsection{\label{ssec2.1}The Weisskopf's theory}
It should initially be reminded that this theory is based on the Bohr's assumption of independence between the phase of formation and de-excitation of the hot nucleus.
 The hot nucleus is assumed to be in thermodynamic equilibrium. The  Weisskopf's theory proposes a statistical description of the process of particle emission by a hot nucleus. For a rather large excitation energy, the nucleus has a continuum of energy states. The transition between states can be described statistically. Weisskopf \cite{weisskopf1} has shown that the emission probability $Prob(\varepsilon) d\varepsilon$  of a particle by an initial hot nucleus, with a kinetic energy between $\varepsilon$ and $\varepsilon +d\varepsilon$, is:
\begin{equation}
Prob(\varepsilon )d\varepsilon=A g p^{2}\sigma _{c}(\varepsilon )\exp (-\frac{\varepsilon }{T}) d\varepsilon
\label{eq1sec1}
\end{equation}
$g$ is the factor of degeneracy due to the spin, $p$ is the particle linear momentum, $ \sigma _ {C} $ is the cross section of capture of the particle by the final nucleus (inverse reaction). $A$ is a factor including different contributions.  It is important to note that the temperature $T$ taken here is the one of the nucleus in the final state therefore the one of the residue. 
 In a classical approximation, where one compares the nucleus to a black body absorbing any particle with its geometrical cross section, and where $B$ is the associated Coulomb barrier, we have then:
\begin{equation}
\sigma_{c}(\varepsilon )=\left\{\begin{array}[c]{c}0\\
\pi R^{2}\left(  1-\frac{B}{\varepsilon}\right)
\end{array}
\begin{array}[c]{c}
si\;\varepsilon \leq B\\
si\;\varepsilon>B
\end{array}
\right. 
\label{eq2sec1}
\end{equation}
The use of this cross section in the equation \ref{eq1sec1}, makes it possible to rewrite this equation in a normalized form ($ \int Prob(\varepsilon ) d\varepsilon =1$):
\begin{equation}
 Prob\left(  \varepsilon\right)  =\left\{
\begin{array}[c]{c}
0\\
C\frac{\left(  \varepsilon-B\right)  }{T^{2}}e^{-\left(  \frac{\varepsilon-B
}{T}\right)  }
\end{array}
\begin{array}[c]{c}
si\;\varepsilon \leq B\\
si\;\varepsilon>B
\end{array}
\right.
\label{eq3sec1}
\end{equation}
$C$ is a constant of normalization.

The equation \ref{eq3sec1} leads to a certain number of interesting properties for this energy distribution. 
We present them below:
\begin{equation}
Prob(\varepsilon ) \: is \: maximum \: at \: \varepsilon=\varepsilon _{max}=B+T 
\label{eq31sec1}
\end{equation}
The mean value of the spectrum is such as:
\begin{equation}
<\varepsilon >=B+2T
\label{eq32sec1}
\end{equation}
The standard deviation of such a function is:
\begin{equation}
\sigma_{\varepsilon}=\sqrt{2} \times T
\label{eq33sec1}
\end{equation}
\subsection{\label{ssec2.2}Presentation of the used experimental method of measurement}
\subsubsection{Principle of measurement}
We saw in the previous section that the Weisskopf's theory makes it possible to foresee the awaited form of energy spectra of the particles evaporated by a hot nucleus. We must be able to reproduce the energy spectrum of a given type of particle by the following function:
\begin {equation}
\frac{dN(\varepsilon )}{d\varepsilon }=C\times \frac{(\varepsilon -B)}{T^{2}}\times exp(-\frac{\varepsilon -B}{T})
\label{eq1sec2}
\end{equation}
The shape of the spectrum depends on three parameters $C$, $B$ and $T$. The two important parameters are the temperature $T$ and the barrier $B$.  We chose to use two fit tests : the $\chi^{2}$ test and the Kolmogorov-Smirnov fit test.
The procedures of adjustment are detailed in the appendix \ref{annexea}. 
 This gives us two estimates of the temperature : $T_ {\chi^{2}} $ and $T_ {Kolmo}$. 
We have also the possibility of determining the temperature by using the equations \ref{eq32sec1} and  \ref{eq33sec1}. 
Indeed, we can deduce from them that
\begin{equation}
T_{\sigma_{\varepsilon}}=\frac{\sigma_{\varepsilon}}{\sqrt{2}}
\label{eq2sec2}
\end{equation}
\begin{equation}
T_{<\varepsilon>}=\frac{<\varepsilon >-B}{2}
\label{eq3sec2}
\end{equation}
For this last one, we need to know the value of $B$. Consequently, we use a value of this variable calculated starting from a weighted average of the values of $B$ obtained by the two fit tests.
\subsubsection{Intrinsic limitations with the method of measurement}
It should be reminded that the temperature which is involved in the equation \ref{eq1sec2} is not the temperature of the initial nucleus but that of the evaporation residue. This theory is logically valid only for the emission of an unique particle. Therefore, the cascade emission, which appears with high excitation energies, is not treated explicitly. It is therefore necessary to suppose a sequential process in the case of the multiple emission of particles.
 After each evaporation, the nucleus loses a part of its excitation energy. It consequently undergoes a progressive cooling.
Moreover the Weisskopf's theory does not treat a possible simultaneous multifragmentation of the nucleus.

The de-excitation moreover generates recoil effects. After each emission, the residual nucleus moves because of the conservation of momentum, which moves it away from its previous position in velocity space. 
However, all calculations of velocity are performed in the initial frame, therefore inducing a systematic error on the determination of the kinetic energy of the following evaporated particles.
We do not know the true order of emission of the particles, so it is impossible to correct this effect in the presence of a sequence of evaporation. 
 
It should also be noticed that there is, a progressive evolution of the emission probability of a given type of particle during the cooling. The mean time of emission of particles so vary according to their nature.  

The particles emitted by the hot nucleus may themselves be excited and therefore decay, giving rise to what is called the "secondary emission".
 These secondary emissions, when they exist, modify the energy spectra, their frame of emission having no relation with the initial frame of the studied hot nucleus. 

There may be also disturbances generated by possible collective effects, such as the expansion or the rotation of the hot nucleus. 

To be capable to apply experimentally this method, it is initially necessary to measure the energy spectrum of the evaporated particles in the frame of the emitting nucleus. There are already, for that, two fundamental difficulties : the reconstruction of the initial velocity of the hot nucleus and the identification of the particles effectively evaporated by this hot nucleus. 

Since measurements are done at the end of the decay chain, the shape of the spectra will reflect on average all these effects. Consequently, the spectra do not present the \textbf{same apparent slope} regardless the nature of the emitted particle, as we will observe thereafter.
There is thus a systematic error due to the manner the temperatures are measured.
One of our goals is therefore to understand the links existing between the \textbf{initial temperature }of the hot nucleus and the \textbf{apparent temperature }associated with each studied particle after the de-excitation chain.
\section{The methodological analysis\label{sec3}}
\subsection{Introduction}
 To simulate the physics of the hot nucleus, we used two statistical models GEMINI and SIMON. The entry parameters used for these two generators are described in the appendix \ref{annexeb}. 
We initially privileged the use of GEMINI. We worked in a range of excitation energy between 0.5 and 4 A.MeV. We chose as reference the nucleus $^ {120} Sn$, because often used by the INDRA collaboration as target.
 To study the influence of the statistics on all our measures, we generated data sets with 1000, 10000 and 100000 events.
 
We voluntarily prohibited the emission of intermediate mass fragments ($Z \ge 5$). The used version of GEMINI treats this emission less correctly \cite{charity2}. 
The level density parameter was assumed constant and equal to $A/10$. 

\begin{figure} [htbp]
\centerline {\includegraphics [width=8.7cm, height=17.52cm] {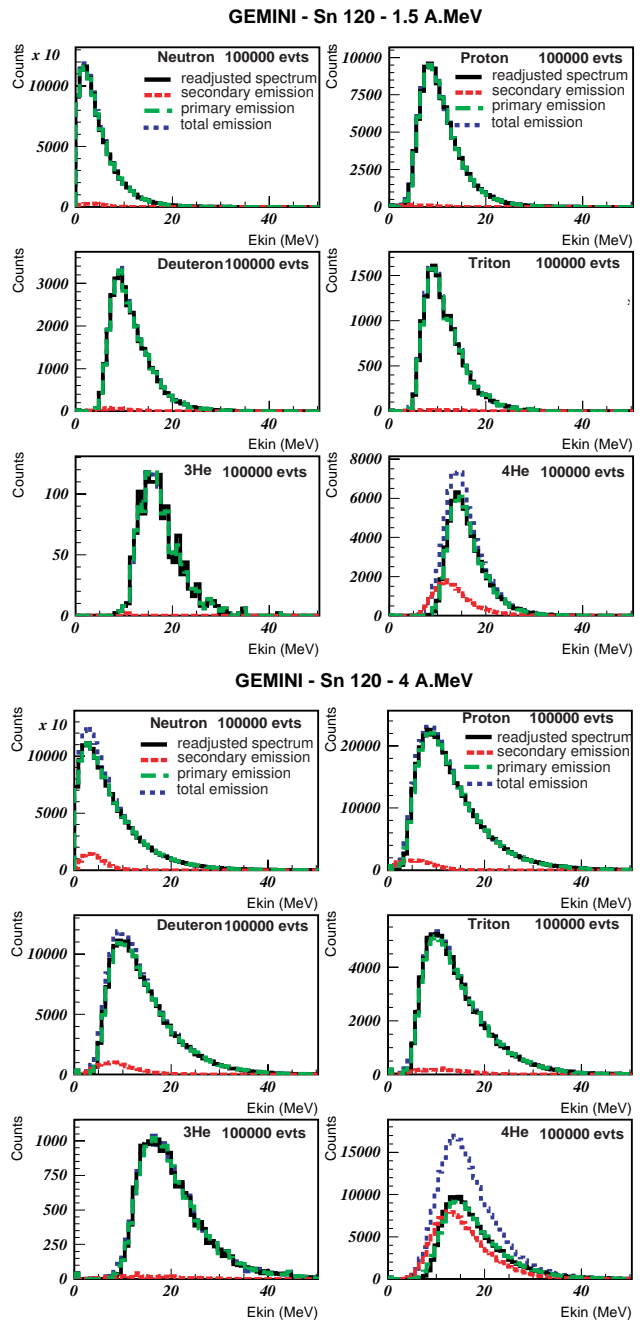}}
\caption {Evaporation energy spectra of light particles for the excitation energies : 1.5 and 4 A.MeV. The blue spectrum (dotted) corresponds to the totality of the emitted particles, the green (dash and dotted) to the primary emission, the red (dash) to the secondary emission and the black (full) is the readjusted spectrum.}
\label{fig1}
\end{figure} 
We will initially present a fast characterization of the evaporation events generated by GEMINI.  We will take the maximal available statistics. 
We present in the figure \ref{fig1} various energy spectra of the particles emitted by $^ {120}$Sn nuclei with an excitation energy per nucleon of 1.5 and 4 A.MeV. These spectra are reconstructed in the initial frame of the hot nucleus (for total emission, primary emission and secondary emission). Since GEMINI can simulate a secondary de-excitation of the unstable excited states of the fragments with a charge lower than 5, it was possible for us to study the respective weights of the primary and secondary evaporative contributions for each type of light particles in the total evaporation spectrum.

 We would  like to observe the influence of the successive recoils of the hot nucleus after each evaporation. We examined also a spectrum, called in the following readjusted, that was deduced after each primary evaporation by  recalculating the new frame of the emitting nucleus and calculated the kinetic energy of the following particle emitted in this frame. The effects of recoil generated by the evaporative cascade seem to be weak at 1.5 A.MeV of excitation energy even for the alphas. As it can be seen in figure \ref{fig1}, for this energy,  secondary evaporation is consequent only for the alphas. At 4 A.MeV, for all the light particles, the total spectra differ from primary spectra. Conversely, the secondary emissions modify the shape of the spectra especially at low energy. For the alphas, the disturbance is much more important. 
The secondary emission represents almost the half of the emitted alphas. At this excitation energy, the difference between the readjusted spectrum and the primary spectrum defined in the frame of the initial nucleus becomes more important, mainly for the most massive particles. The effect appears mainly at low energy.

\begin{figure}[htbp]
\centerline{\includegraphics[width=7.cm,height=7.4cm]{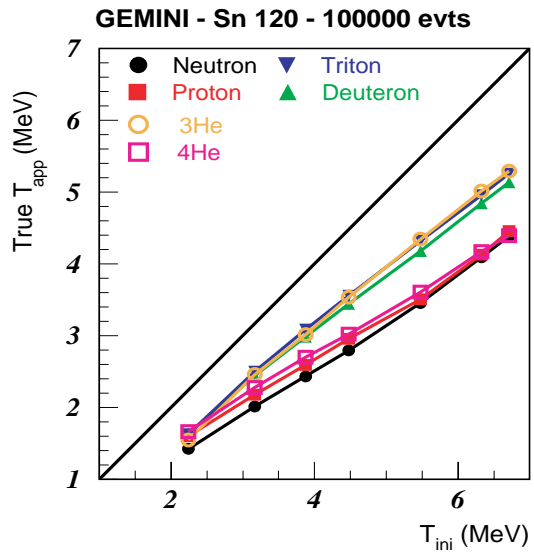}}
\caption{For each type of studied particle, we present the average correlation between the initial temperature of the hot nucleus and the true apparent temperature.}
\label{fig2}
\end{figure}

We have already mentioned above the influence of the cooling of the hot nucleus. Whatever the method, we know that we have a measured temperature, which can only be an apparent temperature since the nucleus temperature varies during the evaporation cascade. However, the use of an event generator allows to monitor the evaporation step by step, therefore the evolution of the nucleus temperature.
We can thus determine the \underline{\textbf{true value}} of this apparent temperature for each type of particle. Indeed, we can calculate the mean temperature felt by each type of particle during a sequence of evaporation. With the used event generator, knowing the excitation energy of the hot nucleus and the level density parameter, we can calculate the initial temperature of the hot nuclei. Therefore, an evolution of these true apparent temperatures according to the initial temperatures of the nucleus can be presented in figure \ref{fig2}. We notice that the curves, associated with each type of particles, are divided into three groups : deuteron, triton and helium 3, then proton and alpha and finally neutron alone mainly for the small initial temperatures. For the first group, the relation between the two temperatures appears to be almost linear.  For the others, there is a clear variation of the slope according to the initial temperature.

\begin{figure} [htbp]
\centerline{\includegraphics [width=6.22cm, height=7.cm] {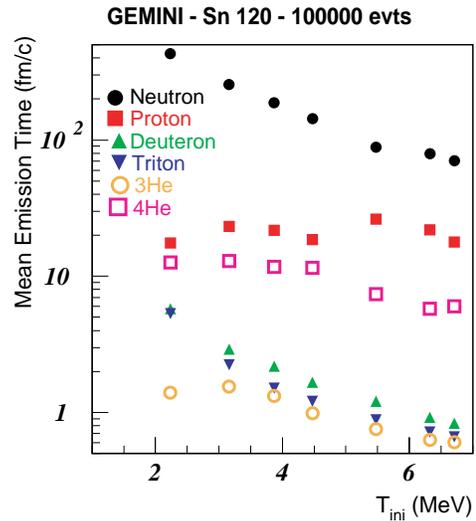}}
\caption{For each type of studied particle, we present the average correlation between the initial temperature of the hot nucleus and the mean emission time of the particle.}
\label{fig3}
\end{figure}

GEMINI provides also the emission time of each evaporated particle. We thus may deduce the mean emission time for each type of particle. We plot on figure \ref{fig3} the evolution of mean emission times according to the initial temperature. We find the groupings already seen in figure \ref{fig2}. The influence of the barriers on the probability of evaporation and 
the life times of the unstable nuclei giving place to a secondary emission, are responsible for this temporal hierarchy. 
 That allows to understand the true apparent temperatures and the differences observed among species. 

We now will try to find these apparent temperatures by applying the thermometry presented previously.

\subsection{"Perfect thermometry" of a "perfectly detected" hot nucleus}

\begin{figure} [htbp]
\centerline{\includegraphics[width=7cm, height=23.cm]{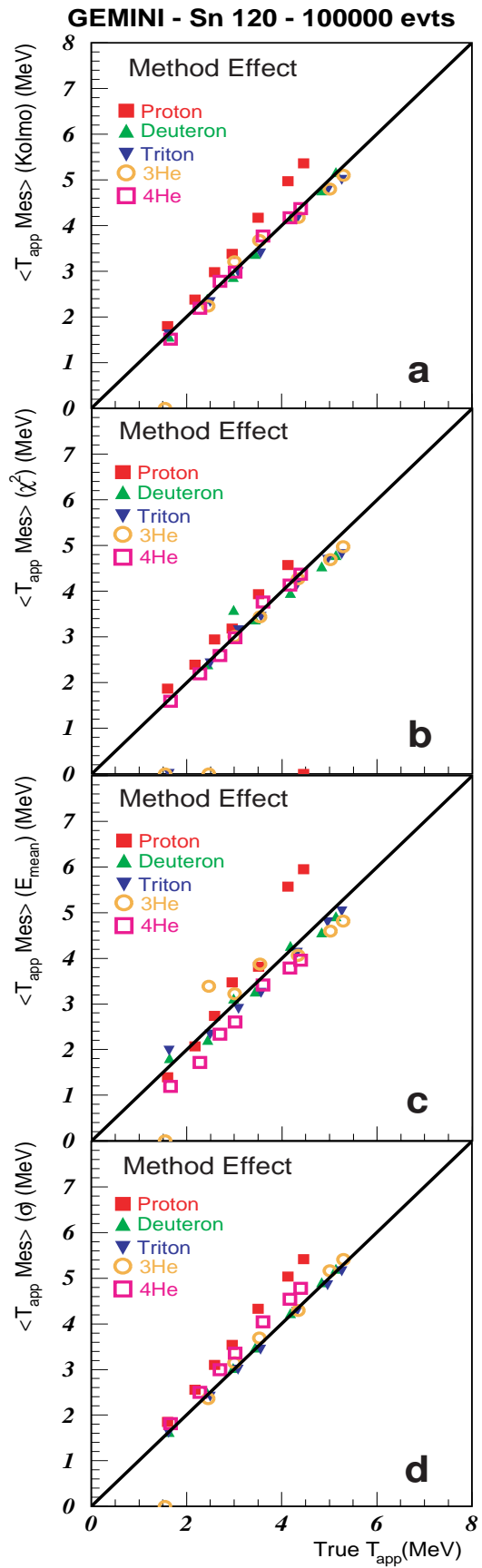}}
\caption{Graphs of the average temperatures, measured from the energy spectra, according to the true apparent temperature for all the light charged particles. These are obtained by the Kolmogorov's method(\textbf{a}), the $\chi^2$ fit (\textbf{b}), from the standard deviation (\textbf{c}) and mean energy (\textbf{d}). The detection and the knowledge of the initial kinematics of the hot nucleus are perfect.}
\label{fig4}
\end{figure}

We initially study the ideal case where we detect all the emitted charged particles and, in addition, the spectra are built in the initial frame of the hot nucleus. Therefore we mention this case as "perfect thermometry". By this type of study, we check the experimental validity of the used thermometry. 
The various methods used to fit  an energy spectrum with a given function, are described in more detail in the appendix \ref{annexea}.
We used  the $\chi^2$ test and the Kolmogorov-Smirnov test in an improved version.
For the thermometry deduced from the standard deviation, we use quite simply the expression given by the equation \ref{eq2sec2}. 
For the measurement of temperature based on mean energy, we apply the relation \ref{eq3sec2} using 
an estimate of the average barrier deduced from the results of the fits. 
But as we do not detect the neutrons with INDRA, thereafter we will consider only the light charged particles. 
We present on figure \ref{fig4} the results of these various measurements according to the true apparent temperature, for all light charged particles and for the maximum available statistics of events.

It seems that the temperature, measured with the help of the $\chi^2$ and K-S fits (see figures \ref{fig4}-\textbf{a} and \ref{fig4}-\textbf{b}), is approximately equal to the true apparent temperature for all the light charged particles except for the protons. For the latter, the measured temperature is systematically above the true apparent temperature. Moreover, the difference increases with it.

Temperature from the standard deviation (equation \ref{eq2sec2}), gives correct results for the deuterons, tritons and helium 3 (see figure \ref{fig4}-\textbf{d}).
On the other hand, the protons and alphas have a systematically over-estimated temperatures. The widening of the observable total spectrum in the figure \ref{fig1}, due to the secondary emissions and the recoil effects, seems to be a possible cause of this trend.
The measures obtained from the average kinetic energy (equation \ref{eq3sec2}) (see figure \ref{fig4}-\textbf{c}), seems more problematic, even if a linear correlation between the measured temperature and the true apparent temperature is found for deuterons and tritons. 
The fact of measuring two quantities like the average energy and the apparent barrier involves an increase of errors, which makes the result more uncertain for the other particles. The estimate of the apparent barrier is difficult as we will see it in the section \ref{sec4} devoted to this measurement.

\begin{figure} [htbp]
\centerline{\includegraphics[width=9cm, height=9cm] {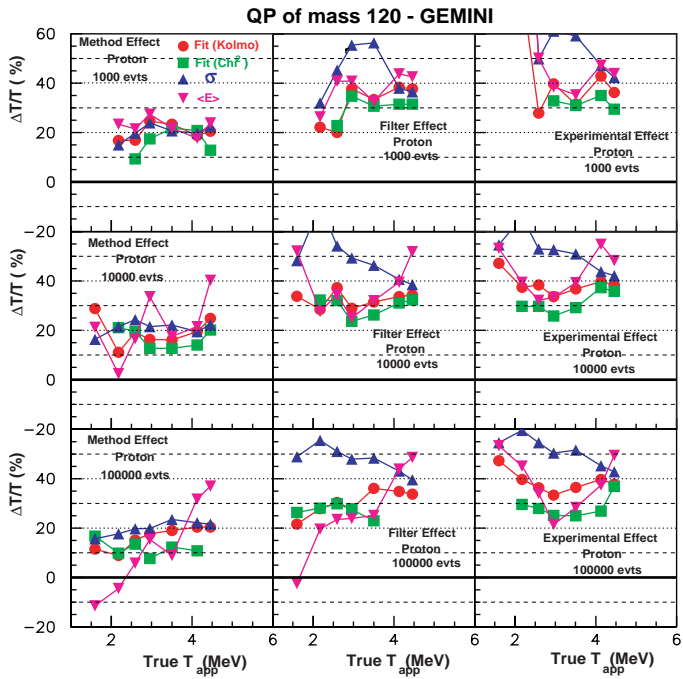}}
\caption{Study of the measures of the apparent temperatures from the energy spectra of protons.}
\label{fig5}
\end{figure}

Figure \ref{fig5} gives quantitative estimates of the agreement between measured apparent temperatures obtained in different conditions and the true value of the apparent temperature given by the event generator and presented in figure \ref{fig4}. In this figure \ref{fig5}, for protons, the relative errors on the measures of the apparent temperature is plotted as a function of the true apparent temperature. Each row corresponds to a given statistics of events. In appendix \ref{annexec} one can find this kind of figure for the others particles.
 At first, for each figure, we will care about the first column, which corresponds to the study of the measurement method, alone. Each column corresponds to one of the experimental situations that we want to study.

For the protons (see figure \ref{fig5}), we find obviously the trend observed in figure \ref{fig4}: the measured temperature is larger than the true apparent temperature, the relative error varies from 20 \% to approximately 15 \%. With a weak statistics, the four techniques are relatively coherent. The methods are more differentiated as the statistics increases.
The measures from standard deviation are generally largest. The $\chi^2$ fit gives a weaker measure than that of Kolmogorov. We will see that the measure of the barrier gives the opposite trend.
The measure deduced from average energy seems more problematic for the largest statistics. That is due mainly to the estimate of the apparent barrier, which is disturbed by the secondary emission, very present at low kinetic energy. The apparent barrier is therefore systematically underestimated for the protons.

For the deuterons (see figure \ref{fig13} in appendix \ref{annexec}), the results appear much better. With low statistics, there are clearly difficulties of adjustment, mainly for  the $\chi^2$ fit. That involves fluctuations on the relative errors that can be as high as 20 \%,  but the average behavior is around 0 \%. That becomes true when the statistics increases.
The measures starting from the standard deviation and of the Kolmogorov's method are even excellent, the relative error being lower than 5 \%.

For the tritons (see figure \ref{fig14}), we can draw the same conclusions as for the deuterons, except the fact that the measured temperature seems slightly lower than the true apparent temperature. 
The relative error is around -5 \%. The coherence between the various methods appears even better than for the deuterons when the statistics is important.

For heliums 3 (see figure \ref{fig15}), there is clearly a difficulty related to the statistics. For 100000 events, the best measures are the ones made from the standard deviation and the Kolmogorov's method. They give between 0 and -5 \% of relative error. 
The $\chi^2$ fit seems less good and converges towards the Kolmogorov's method only for the largest temperatures, therefore for the spectra with the largest statistics.
 
 For the alphas (see figure \ref{fig16}), the methods of adjustment seem to improve with the statistics, whereas the standard deviation and average energy are incorrect whatever the statistics.
 We find even the apparent temperature with less than 5 \% as relative error, what is a little surprising considering the strong secondary emission of alphas. 
This secondary emission, moreover, widens the total spectrum, implying a temperature obtained starting from the standard deviation, which is 10 \% too large. 
We notice also the systematic undervaluation of the temperature obtained starting from average energy, which is also linked to the secondary component of evaporation.

\subsection{Thermometry of a "perfectly known hot nucleus" with INDRA}

We add now solely the effect of the detector array that we call the filter effect.
To be able to filter the events generated by GEMINI, we supposed, for each excitation energy,
 that the hot nucleus had a certain velocity vector in the frame of the laboratory. For each excitation energy, the polar angle between the velocity vector of the hot nucleus and the initial direction of the beam was defined starting from calculations with the generator SIMON,  for collisions Xe + Sn with 50 A.MeV, giving almost the same excitation energy for compound systems of mass close to 120 amu.
 The azimuthal angle was determined randomly. The magnitude of the velocity vector was defined starting from the excitation energy by supposing a total dissipation. 

After having filtered the events, we then reconstructed the energy spectra of the detected light charged particles in the real initial frame of the emitting nucleus. We can analyze them by studying the central columns of the various figures (see figures \ref{fig5}, \ref{fig13}, \ref{fig14}, \ref{fig15} et \ref{fig16}). 
We thus consider the cumulative influences of the measurement method and of the experimental set-up.

For protons (see figure \ref{fig5}), we observe a temperature measured from the standard deviation completely different from the others.
 This difference decreases when the temperature increases. This is true whatever the statistics. Other methods of measurement give also larger temperatures systematically. The relative error increases by approximately 10 \%.
 
 For deuterons and  tritons (see figures \ref{fig13} and \ref{fig14} in appendix \ref{annexec}), there is also an increase, but all measures remain coherent. The effect is stronger for the lowest temperatures especially for the deuterons. It thus must be correlated with the kinematics chosen for the hot nuclei. The relative error increases of 20 \% on average.
 
For helium 3 nuclei (see figure \ref{fig15} in appendix \ref{annexec}), the effect of the filter is extremely important.  Certain method gives relative errors greater than 60 \%. Only for  $\chi^2$ test, with an important statistics, we observe a relative error ranging from 60 \%  to 12\%  when the apparent temperature increases. 

For alphas (see figure \ref{fig16} in appendix \ref{annexec}), there is always a dispersion of the results of the various methods. It is necessary to add  a relative error to that which grows 30 \%  overall. This variation decreases with the true apparent temperature, as for the deuterons.

We can conclude that the experimental device apparently strongly distorts thermometries obtained from energy spectra.
\subsection{Understanding the effect induced by the filter}
To better understand the effect of the filter, we display on table \ref{tab1} for various excitation energies of interest, a percentage of events, on the one hand for which at least one particle was correctly detected by INDRA and on the other hand for which, at least 70 \% of the initial linear momentum and 50 \% of the initial charge, were collected by INDRA. This last selection is called complete events. The SIMON calculations indicate a strong forward focusing of hot nuclei, like in real data \cite{steck1}. 
This involves important difficulties of detection for lowest excitation energies, as table \ref{tab1} clearly suggests it.
The residue goes systematically through the beam hole around zero degrees. Driven by the initial motion of the nucleus, the evaporated particles, with not enough kinetic energy, must be extracted from the hole. At 1.5 A.MeV, at least one charged particle produced in an event is detected, but only 1/3 of the evaporation residues are detected, necessary to obtain a complete event. 
\begin{table}[htbp]
\begin{center}
\begin{tabular}{|p{2.3cm}|p{.8cm}|p{.8cm}|p{.8cm}|p{.8cm}|p{.8cm}|p{.8cm}|}
 \hline
E*/A in A.MeV & 0.5 & 1 & 1.5 & 2 & 3 & 4 \\
 \hline
Filtered evts& 16 \% & 75 \% & 97.2 \% & 99.8 \% & 100 \% & 100 \% \\
 \hline
Filtered and complete evts& 6.3 \% & 23.6 \% & 33.1 \% & 38 \% & 43.9 \% & 49.8 \% \\
 \hline
\end{tabular}
\caption[Detection efficiency of the events]{Study of detection efficiency of the GEMINI events by the detector array INDRA.}
\label{tab1}
\end{center}
\end{table}
 The strong focalization towards the front generates a sampling of particles which is not indifferent concerning the kinetic energy. The energetic spectra are disturbed. We keep only events with evaporated particles having a sufficient kinetic energy, consequently a sufficient linear momentum, to eject the evaporation residue out of the beam hole. This effect has been already observed in real data and is clearly shown and explained in two papers \cite{Vient1, steck1}. It is called the right-left effect.
\begin{figure} [htbp]
\centerline{\includegraphics[width=8.7cm, height=17.3cm] {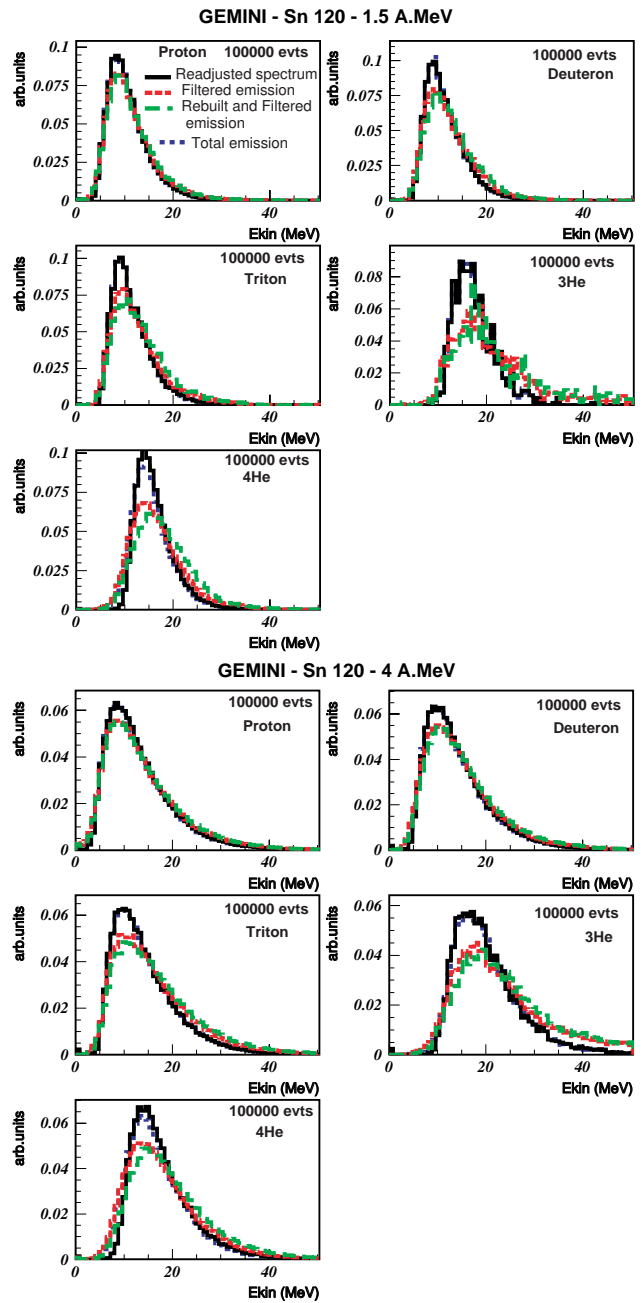}}
\caption{Normalized energy spectra of evaporation of the light particles for excitation energies of 1.5 and 4 A.MeV.
 The blue spectrum (dotted) is defined in the initial frame and corresponds to the totality of the emitted particles. The red spectrum (dash) is defined in the initial frame and corresponds to the emitted and detected particles.
The green spectrum (dash and dotted) is defined in the reconstructed frame for the well detected events. The black spectrum (full) is the ideal spectrum, i.e., primary emission and readjusted after each evaporation.}
\label{fig6}
\end{figure}

In  figure \ref{fig6} are shown same spectra for each type of evaporated particle at two different excitation energies, 1.5 and 4 A.MeV. 
As one can see, figure \ref{fig6} shows two spectra taken as reference  (see previously in figure \ref{fig1}) : the primary emission readjusted spectrum, being "the ideal spectrum", defined in the frame of the emitting nucleus, recomputed after each evaporation, and the spectrum  with all the emitted particles obtained by calculating the kinetic energy in the initial frame of the hot nucleus that is the "ideal experimental spectrum". For these spectra, the detection is perfect as in figure \ref{fig1}.

Two supplementary spectra are also shown : the spectrum of the particles, detected by INDRA, whose kinetic energy is calculated in the initial frame of the emitting nucleus and the one of the same particles, whose kinetic energy is defined in the reconstructed frame. 
All these spectra are normalized to 1. We can observe a systematic widening of the spectra after the filter.
This is more apparent when the particle is heavier. That is obviously the cause of the systematic increase of T$_{\sigma}$.
Due to this fact, the spectra display also weaker apparent slopes and  therefore larger apparent temperatures.

 Hot nuclei, strongly focused forward, as well as the limited angular resolution of  INDRA must be partly responsible for that. Indeed, this geometrical limitation generates an important uncertainty concerning the direction of the velocity vector of the particle.
 This fact plays a role in the calculation of the particle energy in the frame of the initial nucleus. The distortion on this energy is larger as the particle is heavier or is faster.
This forward focalization involves also an important probability of double detection in the same telescope. In this case, there is often a bad identification and an incorrect energy. The most sensitive particles to this type of pollution are those characterized by the smallest multiplicities and being able to be "imitated" by a double counting of the most probable emitted particles. This may be the case for nuclei such helium 3.
 
\begin{figure} [htbp]
\centerline{\includegraphics[width=7.5cm, height=7.5cm] {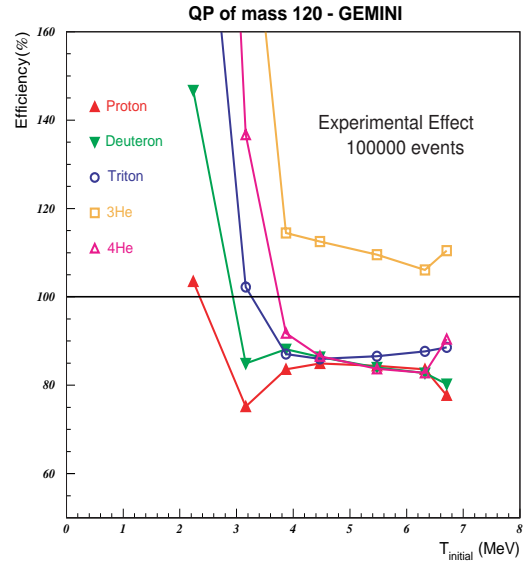}}
\caption{Study of the detection mean efficiency of the various light charged particles.}
\label{fig7}
\end{figure}

To supplement this information, we also plotted in the figure \ref{fig7} the detection efficiency of each type of particle (ratio of mean detected multiplicity for the studied events over the mean multiplicity for all the generated events) according to the initial temperature. We did it for the maximal available statistics and for the complete events.

We can remark two important facts.  First, we have an abnormal apparent efficiency for the weakest initial temperatures for all types of particle. It is associated with the difficulty of completely and correctly detecting the most peripheral collisions. For these collisions, the mean multiplicity is very weak, below unity. Therefore we understand that, mainly the events for which the multiplicity of a type of particle is higher, can more favorably take out the residue of the hole for the beam. This is more evident when the excitation energy is low and the velocity of the evaporation residue is large. For the other collisions, all the particles, with the exception of helium 3, present a roughly constant efficiency around 85 \%. It is the expected efficiency for these ones with INDRA \cite{Pouthas1, Vient2}.
Second, there is clearly an excess of $^3$He nuclei for all temperatures, which can be explained only by a double counting of other particles, more abundant, and which makes it possible to understand the important increase of the high energy contribution of their spectrum, the energies of two detected particles being summed. Obviously, these results are related to the realism of our experimental filter, which tries to take into account most realistically all the phases of the detection up to the experimental method of identification.
\subsection{Thermometry of a hot nucleus with INDRA}
Finally, we examine the experimental situation corresponding to that encountered when we look at real experimental data. For the events called "complete", we study the energetic spectra of the detected particles whose kinetic energy is calculated in the frame,  reconstructed starting from the fragments of charge higher than 3. Thus, we analyze the cumulative effects of the method of measurement,  of the experimental filter and finally of the method of reconstruction of the hot nucleus kinematics. All these effects represent the complete experimental protocol as it is applied to real experimental data.
 All the thermometric measurements, corresponding to this experimental situation, are presented in the last column of each figure associated with each type of particles.
 
For the protons (see figure \ref{fig5}), there is particularly an increase of the relative error at low temperatures of about 15 \%. It tends to disappear when the apparent temperature increases.

For deuterons (see figure \ref{fig13} in appendix \ref{annexec}) there is also an additional but smaller effect at low temperatures. The relative error remains lower than 20 \% for the hottest nuclei. 
The difference between the  various thermometries increases slightly.

For the tritons (see figure \ref{fig14} in appendix \ref{annexec}), there is also an increase at low temperature but there is especially a clear increase of 10 \% at the highest temperatures.
 
Helium 3 nuclei cannot be used (see figure \ref{fig15}).  For  alphas (see figure \ref{fig16}) we observe the same trends as for the other particles, with a total drift up to 10 \% and an effect which is more pronounced at low temperatures.

 \subsection{Understanding the cumulative effect of the filter, the reconstruction procedure and the completeness}
 
 There is clearly a distortion at low temperatures for measurements made starting from the energy spectra. 
The uncertainties related to the reconstruction of the frame of the emitting source increase further the errors in the determination of the particle kinetic energies. These ones always tend to widen the spectrum. We can observe it in figure \ref{fig6}.

It is also necessary to point out the role played by the right-left effect \cite{Vient1, steck1}, when few particles are emitted, which generates a systematic shift between the true and the reconstructed frame of the emitting nucleus. This effect is amplified by the request that events be complete. This criterion selects events for which the heavy residue was detected. This kind of events form a particular sample, namely all events for which the evaporated light charged particles were able to prevent the residue to fall in the beam hole of the experimental set-up.
At low excitation energy, the residue being faster, it is therefore particularly necessary that the light particles are emitted with more energy than normal and often as perpendicular as possible to the initial direction of the hot nucleus.  The effect persists when the excitation energy increases, even if more particles are emitted. 
This explains the systematic shift observed at the right side of figure \ref{fig6} between the detected particle spectra defined in the real initial frame (red curve) and those defined in the reconstructed frame (green curve) for complete events. 
The sampling of particles is not the same one as the table \ref{tab1} and, especially, figure \ref{fig7} indicate. 
Event completeness favors either the events with emitted heavy particles for the less hot nuclei, or events where the multiplicities are more important than normal.
\section{Discussion on the evaporation barriers\label{sec4} }
\begin{figure} [htbp]
\centerline{\includegraphics [width=8.3cm, height=8.3cm] {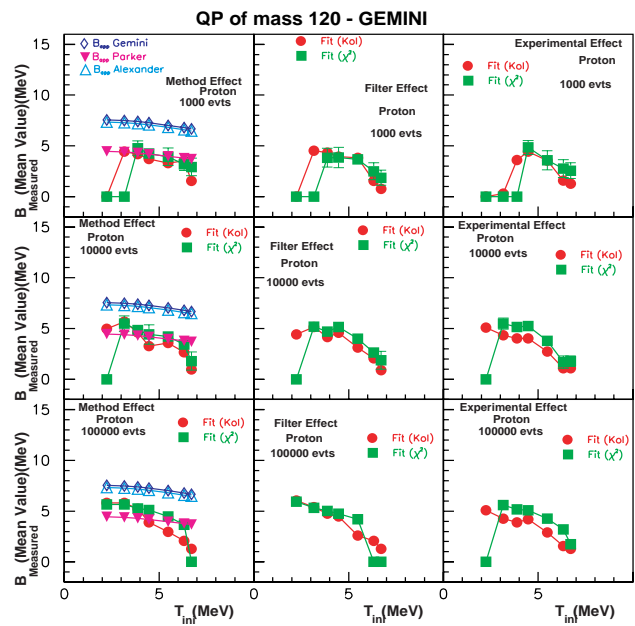}}
\caption{Study of barrier measures from the energy spectra of protons.}
\label{fig8}
\end{figure}

For the measurement techniques, by adjustment of a functional, it appears important to check that the obtained parameter $B$ is also reasonable. This gives us an additional argument to validate the method. We thus present the values of average barriers obtained by the method of the $\chi^{2}$ and that of Kolmogorov-Smirnov as a function of the initial temperature. The followed procedure is the same as the one used for the temperatures. Figure \ref{fig8} concerns the protons.  On the first column  (before applying filter), we chose to present reference barriers for better understanding the quality of the determination of these apparent barriers. These different reference barriers are calculated, firstly  as it is done in GEMINI, secondly by the Parker's formula  \cite{parker1} and by the systematics of  Vaz and Alexander \cite{vaz1}. But, as for the temperature, the evaporation cascade implies a progressive reduction of the size of the emitting nucleus, which involves an evolution of the evaporation barriers. To  account for this effect, we  calculated the apparent barriers starting from those defined in GEMINI, by Parker and Vaz-Alexander. 
We thus determined those by making the average between the initial barrier and the final barrier taking into account the average characteristics of evaporation residues for each excitation energy (see the equation \ref{eq1sec4} below).
\begin {equation}
B_{app} (A_{i}, Z_{i}) = \frac {B (120,50) +B \left (A_{Res} +A_{i}, Z_{Res} +Z_{i} \right)} {2}
\label {eq1sec4}
\end {equation}
It is necessary to point out that the fusion barriers of Vaz-Alexander are empirical barriers obtained from adjustments of experimental fusion cross sections.
Those of Parker are obtained by adjustments of evaporation spectra and angular distributions by using the code of evaporation GANES \cite{ajitanand1} with spherical hot nuclei (angular momentum being also an adjustable parameter).
We will study these figures without doing a specific separation between the various experimental conditions like for the study that concerns  temperatures.

For protons (see figure \ref{fig8}), when the statistics and the multiplicity are low, there are important difficulties to estimate a reasonable apparent barrier by the two methods of adjustment (especially the method of the $\chi^{2}$), this value is often null.
 There is a slight increase of the measured barriers with the statistics.

We notice a systematic difference between the two methods when the nucleus is increasingly hot. 
The Kolmogorov's method gives a barrier slightly weaker than the method of the $\chi^{2}$. 
The measurement of $B$ seems not to be very sensitive to the filter effect and the reconstruction.

For the most important statistics, when we compare the values obtained before filtering with the various values of reference, they appear systematically lower than those expected for fusion or those ones of GEMINI. They decrease relatively in the same way when the initial temperature starts to increase. There is a systematic difference of about 2 MeV. On the other hand, meanwhile, they are larger than those ones of Parker for the weak temperatures. They then seem to decrease and even move under the systematics of Parker for the higher temperatures.
\begin{figure} [htbp]
\centerline{\includegraphics [width=8.3cm, height=8.3cm] {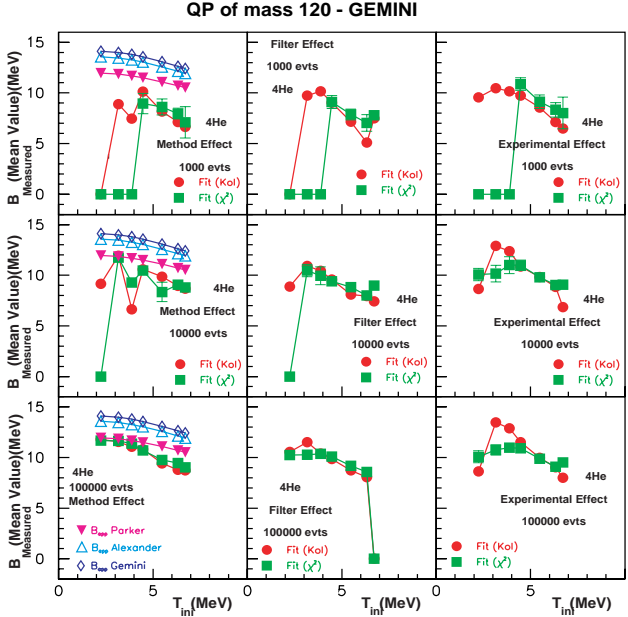}}
\caption{Study of barrier measures from the energy spectra of alphas.}
\label{fig9}
\end{figure}

For alphas (see figure \ref{fig9}), we find the difficulties observed with the other particles
 for the spectra presenting a weak statistics. The measured barrier seems to increase slightly when statistics increases.
 The filter seems to have a weak effect; on the other hand, the barriers measured by the Kolmogorov's method for weakest excitation energies, increase a little when we reconstruct the frame of the emitting nucleus and select complete events.
For the largest statistics, the measured barriers are between 2 and 3 MeV lower than those ones of GEMINI or of the systematics of Vaz-Alexander \cite{vaz1}.
 The difference increases with the excitation energy of the nucleus.
Initially equal at low temperature with those ones of Parker, they become lower by almost 2 MeV for the largest temperatures.

\section{Conclusions on this analysis\label{sec5}.}
We studied the possibility of measuring the temperature of $^{120}Sn$ nuclei starting from spectra of light charged particles. 
We built on purpose these spectra starting from a sequential de-excitation of $ ^{120}Sn $ nuclei. This task has been accomplished only by evaporation of light particles. 
It is the type of disintegration which seems the most consistent with the theory on which are mainly based the studied thermometers.
We used for that two event generators, GEMINI between 0.5 and 4.5 A.MeV  and SIMON between 0.5 and 7 A.MeV. We presented in this paper only results obtained with GEMINI. The two simulations being used under equivalent physical conditions, it appears that the results and the conclusions of this study are independent of the used generator. 

By a quite specific protocol, we were able to deconvolute  the various physical and experimental effects acting on the estimate of the temperature.

We tested two different methods of measurements : first, the fit by the method of the $\chi^{2}$ or of Kolmogorov; second,  from the measure of the standard deviation $\sigma$ of the kinetic energy distribution.
For important statistics, concerning deuterons and the tritons studied by a perfect detector, we built energetic spectra, defined in the true frame of the initial hot nucleus. In this case, the two methods of fits allowed us to measure the true apparent temperature associated with these particles with an accuracy better than 10 \%.
On the other hand, in the same situation for the protons, the measures are rather between 10 and 20 \%. They are less good when one uses the standard deviation of the spectrum.  
The fact that the protons are emitted in larger quantity and on a longer range of time, disturbs the possible correlations between the spectrum and the true apparent temperature.
For the alphas, the apparent temperature is measured with an accuracy better than 10 \% even if the spectrum is strongly disturbed by the secondary emission at low energy.
Clearly for the alphas, this phenomenon prevents the temperature measurement from the standard deviation.

Even if the used kinematics can be discussed, it was clearly demonstrated that the detection set-up plays a fundamental role on the quality of the temperature measurement.
 
 After the experimental filter, measures appear clearly distorted and doubtful. In the most favorable case, i.e. for deuterons and tritons, we observe from 10 to 30 \% approximately of errors for deuterons and 20 to 30 \% for tritons.
The particle detection accounting for kinematics and for the angular resolution of the detector implies an obvious distortion of the spectra for all types of particle. For deuterons and the tritons, the effect is dominant for the weakest excitation energies and decreases a little for the largest ones. The different methods of measurement are consistent from this point of view. For protons and alphas, these effects are much more important and the temperature measurement from the standard deviation appears inaccurate.

The reconstruction of the initial frame and the condition of event completeness disturb only very few the measurement in addition. They do it mainly for the weakest excitation energies. This is perhaps related to the choice of prohibiting the emission of fragments of too important size, which therefore favors an emission of more energetic light particles.

We also pointed out the importance of the statistics for this kind of study.
We saw that the methods of adjustment cannot be used without thinking. The quality of the adjustment and the values measured depend on actual statistics in the spectrum (number of studied events and multiplicity).
The method of the $\chi^{2}$ requires a minimal statistics. The Kolmogorov's method can be used for weaker statistics. When statistics increases considerably, it is simpler to use the method of the $\chi^{2}$ because the criteria of convergence are less drastic. The adjustment is done with a more homogeneous manner for the entire spectrum.
We can notice that the measured apparent temperature seems to increase slightly when the statistics decreases.

By the methods of adjustment, the measured apparent barriers appear completely coherent with respect to the measured apparent temperatures, when these ones are correctly measured. They are systematically lower than expected. This effect is more important when the evaporative secondary contribution becomes important. This result is a little surprising, knowing that there are no angular momentum nor deformation of the hot nucleus. It appears not to be coherent with respect to the results presented by R.Charity in the reference \cite{charity3}.
But this could be related to the choice to impose a de-excitation only by light particle emission in GEMINI. Indeed, it favors the emission of excited light particles and thus increases secondary de-excitation.
On the other hand, the estimate of the barriers is disturbed apparently little by the influence of the filter 
or by the experimental method of reconstruction of the energy spectra.

We can deduce from this study that it is possible to measure the true apparent temperature from the energy spectrum of light particles, if those are produced very early in the most limited possible range of time and few during secondary de-excitation. It is indeed the case of deuterons and tritons, for GEMINI and SIMON such that we used them.
We need obviously the most important possible statistics. In experiments, we need an angular coverage and an angular resolution better than INDRA. It is necessary as much as possible to limit the size of the hole which lets pass the beam and ideally to use a magnetic spectrometer for the very peripheral collisions.
 We voluntarily neglected all the collective phenomena, like expansion or rotation, or even the processes of de-excitation, which could move us away from the ideal conditions of applications of our spectral thermometries. They will obviously not improve these measurements.
 
To finish and complete this analysis, we also made an additional study not presented here, following the same methodology, concerning both standard methods proposed in the literature by  Wada \cite{wada1} and Hagel \cite{hagel1} to trace back to the spectra of first emission and consequently almost to the initial temperature of the hot nucleus. We noticed that the first method does not allow to find the initial temperature of the hot nucleus. In the most favorable situation, corresponding to a perfect detection and a statistics of 100000 events, this one is systematically underestimated whatever is the method of measurement and the type of particles.  When the influences of the INDRA filter and of the reconstruction method of the frame of the emitting nucleus are added, the obtained values are overestimated in an important way. The second considered method allows to determine the initial temperatures to better than 15 \% for deutons, 10 \% for tritons, helium 3 and alphas by using the Kolmogorov's method within the framework of a perfect detection and of a statistics of 100000 events. In other situations, it does not seem more useful.
\appendix
\section{Function adjustments on energetic spectra\label{annexea}}
\subsection{Introduction \label{section1annexea}}
We chose to use two different methods of adjustment to estimate the temperature of the hot nucleus : the classical  $\chi^{2}$ method and the Kolmogorov-Smirnov s' method improved by Kuiper \cite{hpress1}. 
This choice is due to our will to study the influence of the statistics on the quality of our measure. 
Indeed the Kolmogorov's method is very powerful when the statistics becomes weak. These two statistical methods are also of a great interest because it is possible to calculate a probability of compatibility of the experimental distribution with the theoretical distribution. We can thus have a quantitative criterion allowing to judge the quality of the adjustment.
\subsection{Presentation of the two methods \label{section2annexea}}
\subsubsection{Test of the $ \chi^ {2}$}
Let an experimental distribution $N_ {i} (X)$ of a physical variable $X$, in the form of a histogram of $k$ channels, built from $n$ independent measurements. 
It is supposed that the distribution of this variable $X$ obeys a certain law of probability $P(X)$ dependent on $p$ known parameters or to determine.
The difference between the experimental sample and the expected theoretical distribution can be measured by what is called the distance from the $ \chi ^ {2} $, $D_ {\chi ^ {2}} $ defined in the following way :
\begin{equation}
D_ {\chi^{2}} = \sum_ {i=1}^{k}\frac {(nP_ {i} (X) - N_{i} (X)) ^ {2}} {nP_ {i} (X)}
\end{equation}
Under the assumption $H_ {0} $ that the theoretical law of distribution $P (X) $ is indeed that one to which the variable $X$ obeys, the variation $nP (X_ {i}) - N_{i} (X_ {i}) $ between the theoretical distribution and the experimental distribution is distributed according to a normal statistical law.
Under these conditions, $D_ {\chi ^ {2}} $ tends towards a $ \chi^{2} $ law with $\nu$ degrees of freedom ($ \nu= k  - 1 - p $). The acceptance region of the $\chi^{2}$ test, therefore of the assumption $H_ {0} $, is the interval $ (0, \chi_ {\nu, 1 - \alpha} $)
 such that the probability that a variable $\chi^{2}$ takes a value in this interval, is equal to $1- \alpha $.
$ \alpha $ is called the significance level. It is said that the assumption $H_ {0}$ is accepted with the significance level $ \alpha $, if $D_ {\chi^{2}}$ is lower than $\chi_{\nu, 1-\alpha}$.
 
Taking into account that, we can define an indicator $Proba_ {\chi^{2}} $ \cite{hpress1}, called p-value, which makes it possible to quantify the quality of the adjustment.
It is calculated in the following way:

\begin{eqnarray}
Proba_{\chi ^ {2}} =1- \alpha (D_ {\chi^{2}}, \nu) =1-f(\frac {D_ {\chi^{2}}} {2}, \frac {\nu} {2} )\nonumber
\end{eqnarray}
with
\begin{eqnarray}
f(\frac {D_{\chi ^ {2}}} {2},\frac {\nu}{2}) =\frac {1}{\Gamma \left (\frac {D_{\chi^{2}}} {2} \right)} \times \int_ {0}^{\nu/2}e^{-t} \times t^{\frac {D_ {\chi^{2}}} {2} - 1} dt
\end{eqnarray}

The closer $Proba_ {\chi^{2}}$ is to 1, the more the assumption $H_{0}$ appears probable. 
 
\subsection{ Kolmogorov-Smirnov's test}

The problematic is the same one as previously above. But for this test, we compare now the
 empirical function of distribution $F_{i}(X) $ and the theoretical function of distribution $F_{i}^{\ast}(X)$, that we build in the following way:
\begin{eqnarray}
 F_{i}(X) = \frac {1} {n} \sum_{j=1} ^ {i} N_{j}(X) \nonumber
\end{eqnarray}
and
\begin{eqnarray}
F_{i}^{\ast} (X) = \sum_{j=1}^{i} P_{j}(X)
\end{eqnarray}
Here, $i$ ranges from 1 to $k$.\\
The difference between the two distributions is given by the absolute value $D_{+}$ of the maximum distance between $F_{i} (X) $ and $F_{i}^ {\ast}(X)$, called Kolmogorov-Smirnov's distance and defined in the following way:
\begin{eqnarray}
 D_{+} =\underset {- \infty<X<+ \infty} {\max}|F_{i}(X) - F_{i}^{\ast}(X)|
\end{eqnarray}
The Kolmogorov's test has the defect to have a sensitivity which depends on $X$.  
To limit this effect, there exist alternatives: the statistics of Anderson-Darling or that of Kuiper \cite{hpress1}.
We chose that of Kuiper. We thus calculate now a new distance $V$, which is the maximum sum of the distances when, on the one hand $F_{i}(X)$ is above $F_{i}^{\ast}(X)$, and on the other hand the opposite :
\begin{eqnarray}
V= D_{+} + D_{-} 
\end{eqnarray}
with 
\begin{eqnarray}
D_{-}=  \underset {- \infty<X<+ \infty} {\max}|F_ {i} ^ {\ast}(X) - F_{i}(X)|\nonumber
\end{eqnarray}
It is possible under this statistics also to define an indicator to quantify the quality of the adjustment.
This is defined as follows:
\begin{eqnarray}
Proba_ {KP} =Q_ {KP} \left( \left[ \sqrt{N_ {e}}+0,155+\frac{0,24}{\sqrt{N_ {e}}}\right]  \times V\right) 
\end{eqnarray}
with 
\begin{eqnarray}
Q_{KP} (\lambda)=2 \sum_ {j=1} ^ {\infty} (4j^ {2} \lambda^ {2} -1) e^ {-2j^ {2} \lambda^ {2}} \\
 N_{e} = \sum_ {i=1} ^ {k} N_ {i} (X)=n
\end{eqnarray}

\subsection{Applications of these methods to the temperature measurement \label{section3annexea}}

From a practical point of view, we chose to build energy spectra of 250 channels (1 channel having a width in energy of 1 MeV). 
We decided to adjust only spectra having statistics higher than 150 counts. The energy spectra not exceeding 60 MeV, that leaves reasonable statistics by channel for a major part of the spectrum (a necessary limit of 5 counts per channel is often evoked for the  $\chi^2$ test).

\begin{figure} [htbp]
\centerline{\includegraphics[width=8.cm, height=8.cm] {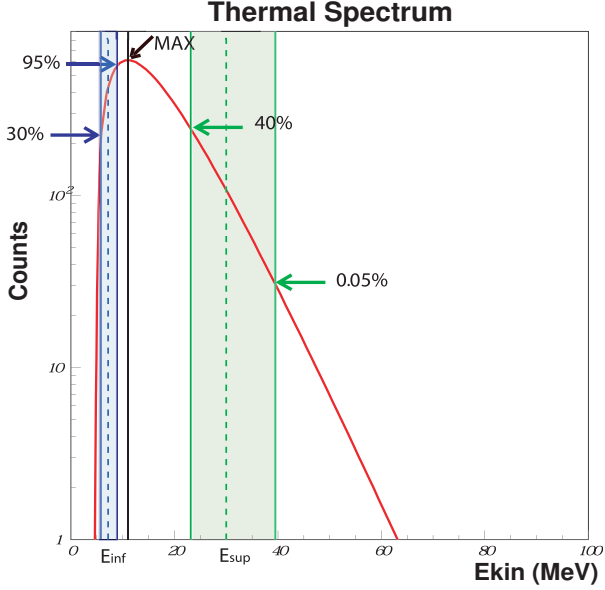}}
\caption{Presentation of the various zones of variation of the limiting values $E_ {inf} $ et $E_ {sup} $ defined within the framework of the $\chi^2$ test.}
\label{fig10}
\end{figure}

\begin{figure} [htbp]
\centerline{\includegraphics[width=8.7cm, height=15.5cm] {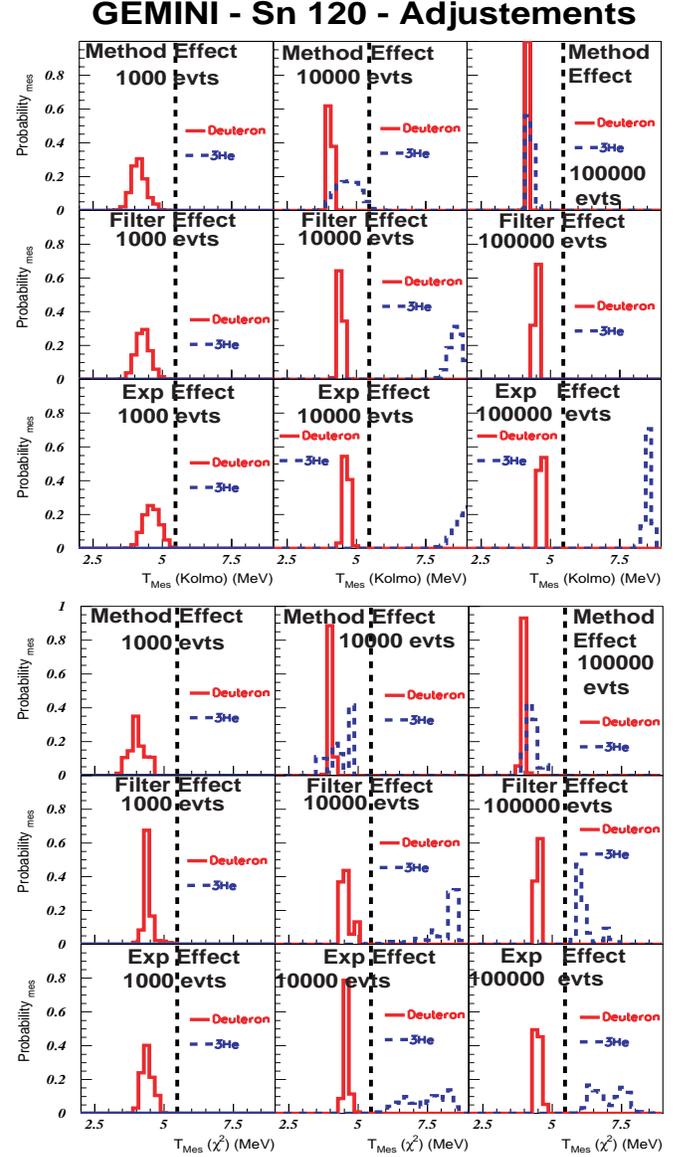}}
\caption{Normalized weighted distributions of the temperatures obtained for the Kolmogorov-Kuiper's test (at the top) and for the $\chi ^2$ test  (at the bottom), for the deuterons, heliums 3 and the different experimental conditions taken into account. The dotted vertical line indicates the value of the initial temperature in the case of a hot nucleus of $^{120}Sn$, having 3 A.MeV of excitation energy per nucleon.}
\label{fig11}
\end{figure}

\begin{figure} [htbp]
\centerline{\includegraphics [width=8.7cm, height=4.6cm] {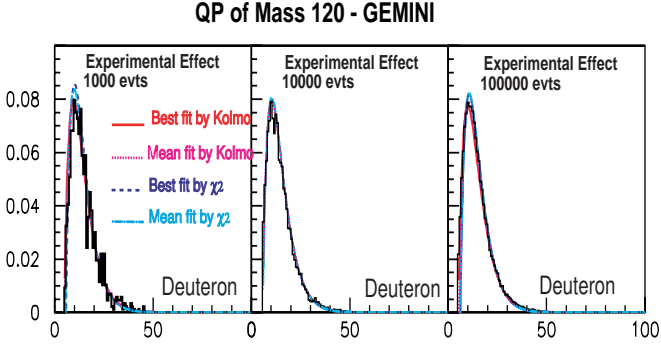}}
\caption{Presentation of the adjustments of the energy spectra of deuteron obtained under the experimental conditions. They are arranged moreover as a function of the number of studied events.}
\label{fig12}
\end{figure}

Classically, the energy spectra are often adjusted by using only a part of the spectrum, to correct the influence of a possible pollution by particles of pre-equilibrium or particles evaporated by another hot nucleus. We therefore chose to adjust each spectrum between two limiting values $E_{inf}$ and $E_{sup}$. We voluntarily varied these two limits to observe the influence of this choice.
 Their zones of variation, defined for the method of the $\chi^2$, are presented in the  figure \ref{fig10}. The limits of the zones are given starting from a percentage of the maximum of the spectrum. The taken zone for the variations of $E_{inf}$ is the same one for the two tests. On the other hand, we took a different percentage for the beginning of the zone of variation of $E_{sup}$ when the Kolmogorov's method is applied: 1 \% instead of 40 \%. This one is less sensitive to the actual statistics by channel.
For each set of values, we sought a set of parameters $C_{\chi^2} $, $B_ {\chi^2} $ and $T_ {\chi^2} $, such as $ D_ {\chi^{2}} $ is smallest possible and another set of parameters $C_ {KP} $, $B_ {KP} $ and $T_ {KP} $, such as the indicator $Proba_ {KP} $ is largest possible.
From that, after variations of $E_ {inf} $ and $E_ {sup}$ in the chosen ranges, we considered two techniques, to determine the apparent temperature and the apparent barrier:
 \begin{itemize}
\item 
First, we took into account the sets of parameters which give the best distance $D_ {\chi^{2}} $ and the best index $Proba_ {KP}$.
\item
Second, we built spectra of parameters $B_{\chi^2} $, $T_{\chi^2} $, $B_{KP} $ and $T_{KP} $ by incrementing those with the values obtained during each made adjustment (for each value of $E_ {inf} $ and $E_ {sup} $), weighted respectively by $Proba_{\chi^2}$ and $Proba_{KP}$. The mean value obtained, for each one of these spectra, is the measured apparent value. These spectra of temperatures are presented in the figure \ref{fig11} for the deuterons and the heliums 3.
\end{itemize}
As an example, are presented in the figure \ref{fig12}, the typical results of  various adjustments for energy spectra of deuteron, emitted by excited nuclei of $^{120}Sn$ with 3 A.MeV of excitation energy per nucleon. 

From these last figures, it is possible to draw a certain number of lessons on these procedures of adjustment.
It is clear that the test of the $\chi^2$ always does not converge when the statistics is weak.
There are no results for the helium 3 nuclei, which have a smaller multiplicity, for 1000 events with 3 A.MeV for example. 
The adjustments are more sensitive to the selected range of adjustment ($E_ {inf} $, $E_ {sup} $) when the statistics is weak. 
Indeed, the values of temperatures obtained are increasingly coherent when the statistics increases (reduction of the standard deviation in the figure \ref{fig11}).

The two measurement techniques (better adjustment and average) in general give equivalent results for the test of Kolmogorov.
The agreement between the two techniques is a little less good with the test of the $\chi^2$. It gets better still when the statistics increases. 	

Now, if we compare the two tests of adjustment, their agreement is all the more correct as the statistics is important. On the other hand, their compatibility seems to depend a bit on the conditions of construction of the spectra, therefore on the shape of the spectra. 
The agreement is less good when the adjusted spectra are those obtained under the conditions of the experiment, being therefore the more different from the ideal spectrum. 

It should be also noticed that the indicator of quality of the adjustment tends to decrease with the statistics of events. Fluctuations decreasing with the statistics, the convergence between the experimental distribution and the theoretical distribution should be better.  
It thus shows us that the convolution of successive de-excitation of the hot nucleus does not give exactly an energy distribution of light particles equivalent to that foreseen by Weisskopf for a single de-excitation. The discrepancy increases even more when are added the experimental distortions.
\section{Description of the conditions of use of the event generators \label{annexeb}}
\subsection {Use of GEMINI \label {section1annexeb}}
GEMINI \cite{charity1, charity2, nicolis1} was used to simulate the sequential emission of a hot nucleus. We imposed that the hot nucleus is spherical with a negligible angular momentum. Angles and the velocity vectors of the decay products are calculated by following a semi-classical mathematical treatment 
to save computing time. This one took into account partly the angular momentum, when it exists. 
We chose not to authorize the emission of fragments of intermediate mass. 
Only particles of charge lower or equal to 5 were emitted. On the other hand, it was taken account 
of their possible secondary de-excitation. The process of evaporation of particles was treated by following the formalism of Hauser-Feschbach \cite{hauser1}.
The calculation of the transmission factors intervening in the estimate of the cross section of the inverse process of evaporation was made within the framework of the model known as (IWBC) suggested by J.Alexander and M.T.Magda \cite{alexander1}.  The used level-density parameter is equal to A/10.
\subsection {Use of SIMON \label {section2annexeb}}
We used the code SIMON \cite{durand1, nguyen1} by imposing a single source. We chose to have a decay of the hot nucleus by sequential evaporation. The Coulomb propagation of the evaporated particles is followed up to 25000 fm/c.
We have frozen the emission of fragments or fission during an extremely long time to emit only the first twenty particles of the mass table used in SIMON.
The emitting nucleus is always of spherical form without notable angular momentum. 
The widths for the various possible ways of decay are calculated by using two different formalisms 
either the theory of Weisskopf \cite{weisskopf1} or the theory of the transition state \cite{bohr2, kramers1}. The latter is used for the emission of particles of size more important than the carbon 12.
We chose to use an option which allows to normalize the widths obtained by Weisskopf by taking account of the widths provided by the theory of the transitory state. 
This makes it possible to ensure a continuity of the emission between light particles and heavier fragments. The secondary de-excitation of the excited light fragments is taken into account. The used level-density parameter is equal to A/10.
The conditions of use of SIMON appears completely equivalent to those taken for the use of GEMINI.
\section{Complementary figures on the study of thermometry\label{annexec}}
 \begin{figure}[!h]
\centerline{\includegraphics[width=8.9cm,height=8.9cm]{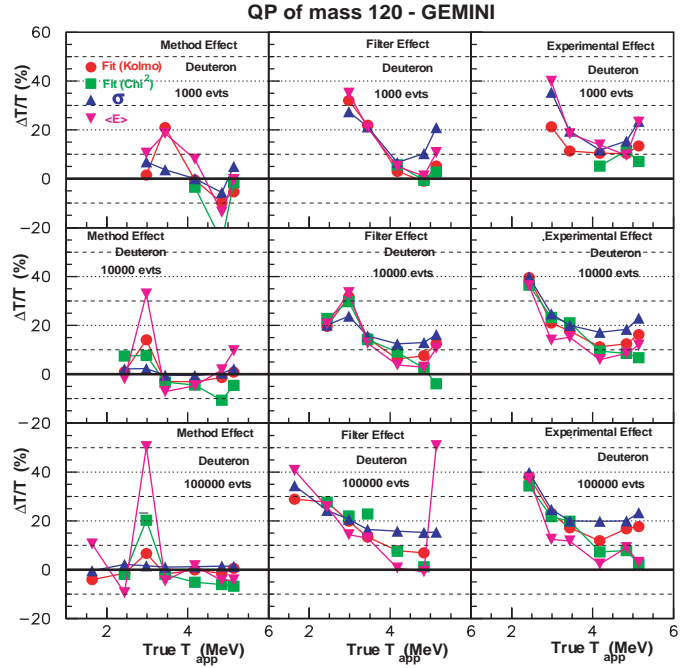}}
\caption{Study of measures of the apparent temperatures from the deuteron energy spectra.}
\label{fig13}
\end{figure}
\begin{figure}[!t]
\centerline{\includegraphics[width=8.9cm,height=8.9cm]{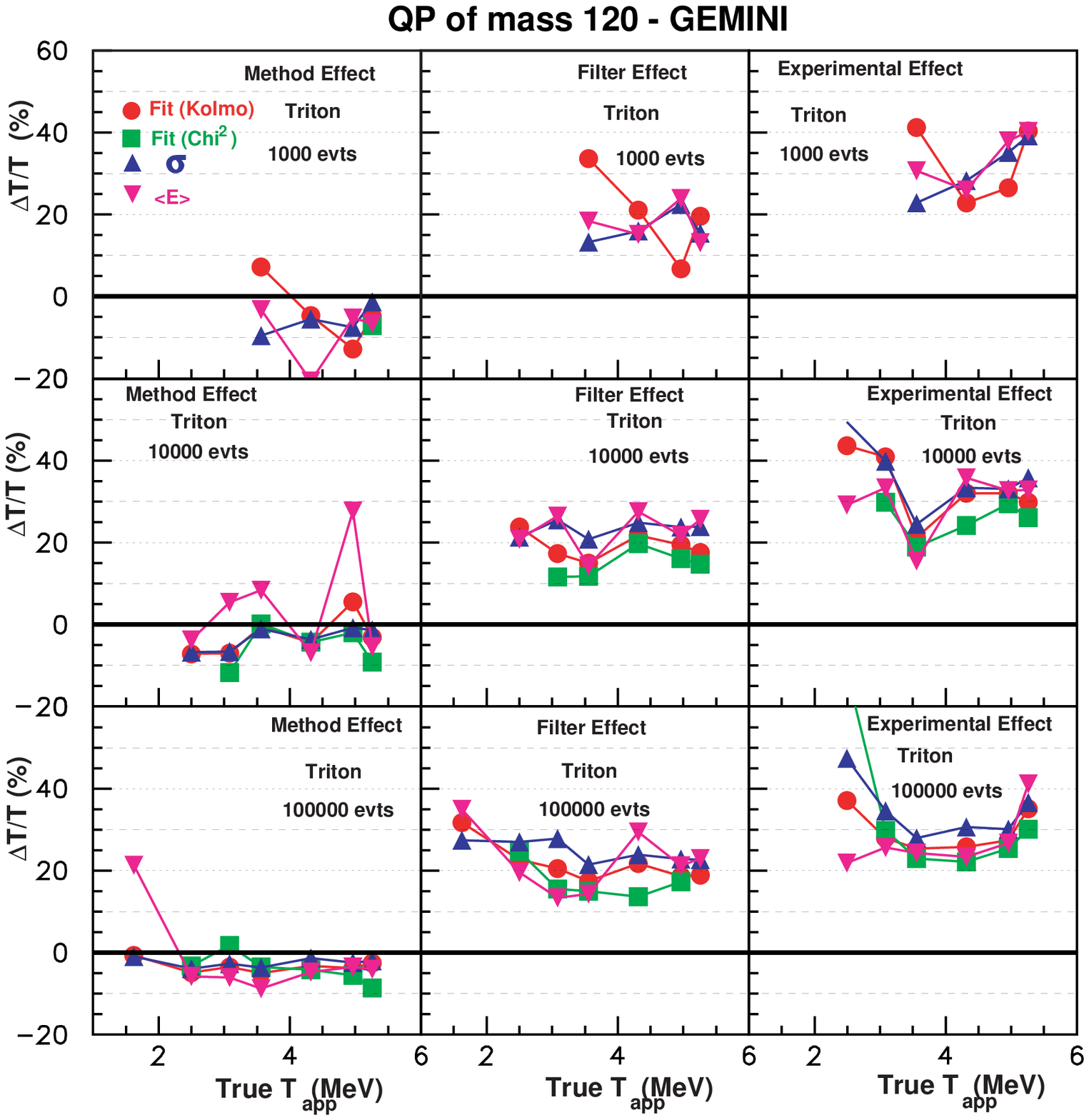}}
\caption{Study of measures of the apparent temperatures from the triton energy spectra.}
\label{fig14}
\end{figure}
\begin{figure}[!b]
\centerline{\includegraphics[width=8.9cm,height=8.9cm]{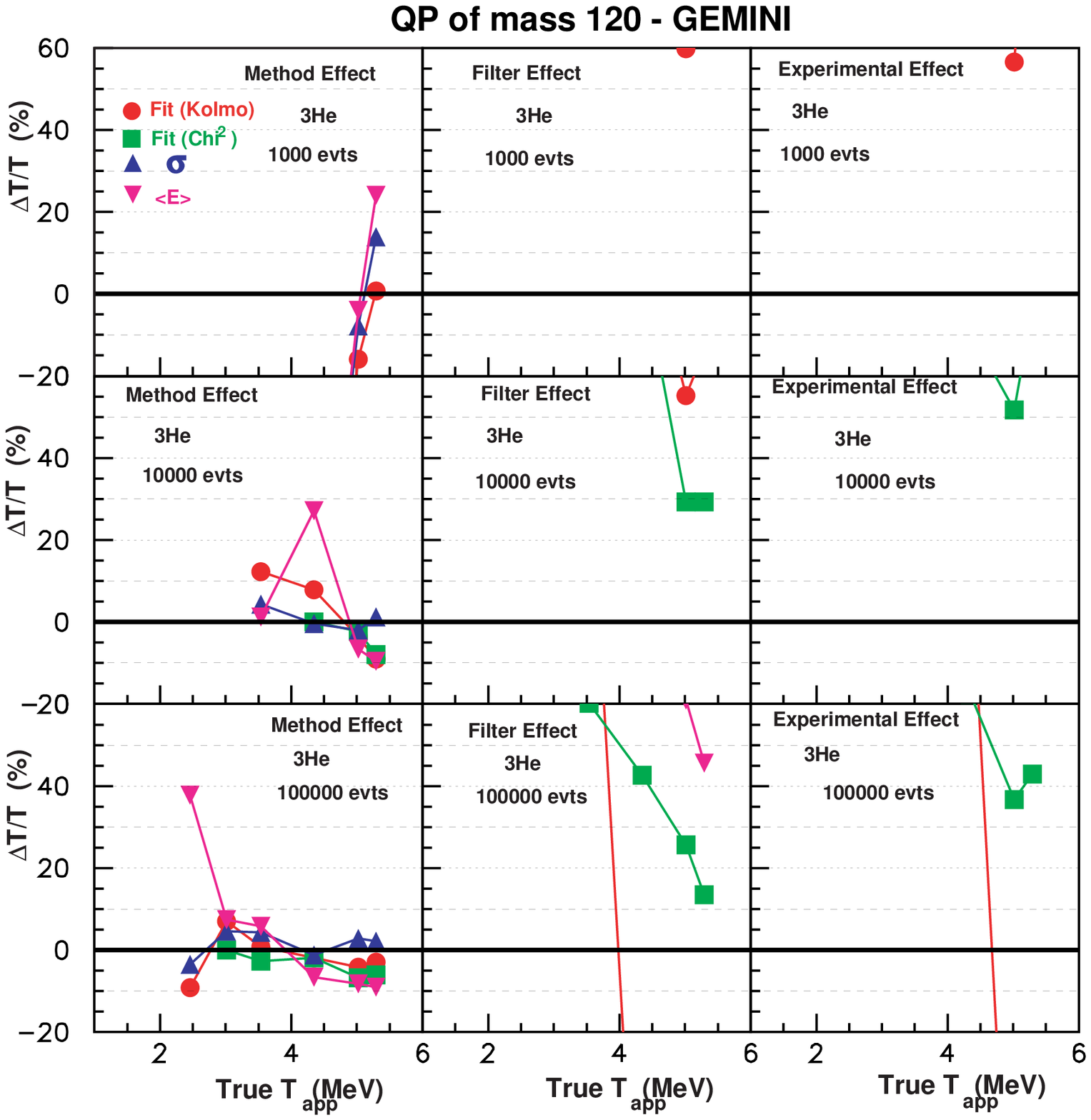}}
\caption{Study of measures of the apparent temperatures from the $^{3}$He energy spectra.}
\label{fig15}
\end{figure}
\begin{figure}[!htb]
\centerline{\includegraphics[width=8.9cm,height=8.9cm]{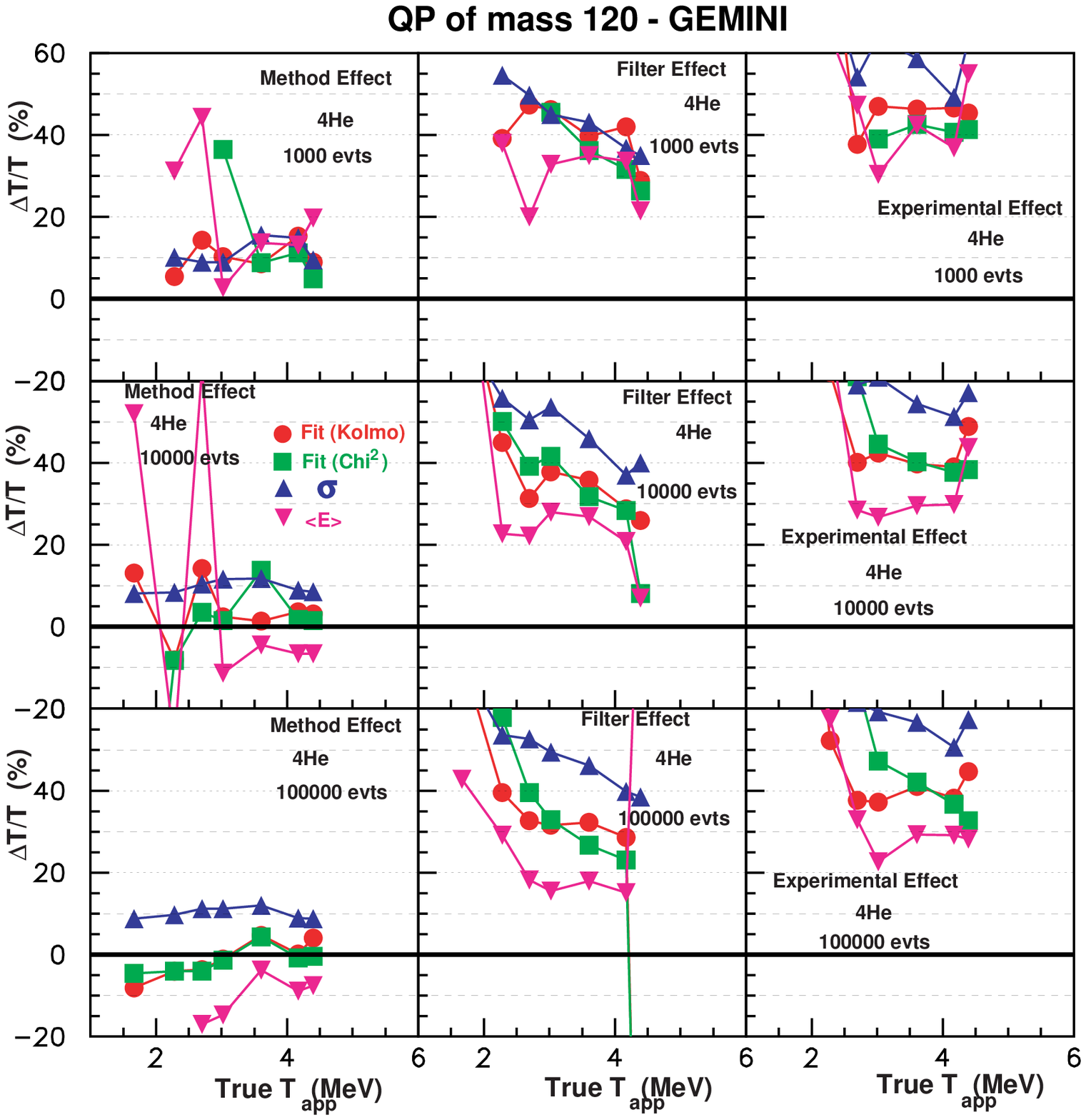}}
\caption{Study of measures of the apparent temperatures from the $^{4}$He energy spectra. }
\label{fig16}
\end{figure}
\pagebreak 
\bibliography{thermometry}

\begin{thebibliography}{41}
\expandafter\ifx\csname natexlab\endcsname\relax\def\natexlab#1{#1}\fi
\expandafter\ifx\csname bibnamefont\endcsname\relax
  \def\bibnamefont#1{#1}\fi
\expandafter\ifx\csname bibfnamefont\endcsname\relax
  \def\bibfnamefont#1{#1}\fi
\expandafter\ifx\csname citenamefont\endcsname\relax
  \def\citenamefont#1{#1}\fi
\expandafter\ifx\csname url\endcsname\relax
  \def\url#1{\texttt{#1}}\fi
\expandafter\ifx\csname urlprefix\endcsname\relax\def\urlprefix{URL }\fi
\providecommand{\bibinfo}[2]{#2}
\providecommand{\eprint}[2][]{\url{#2}}

\bibitem[{\citenamefont{Weisskopf}(1937)}]{weisskopf1}
\bibinfo{author}{\bibfnamefont{V.}~\bibnamefont{Weisskopf}},
  \bibinfo{journal}{Phys. Rev.} \textbf{\bibinfo{volume}{52}},
  \bibinfo{pages}{295} (\bibinfo{year}{1937}).

\bibitem[{\citenamefont{Ericson}(1960)}]{ericson1}
\bibinfo{author}{\bibfnamefont{T.}~\bibnamefont{Ericson}},
  \bibinfo{journal}{Advances in Phys.} \textbf{\bibinfo{volume}{9}},
  \bibinfo{pages}{425} (\bibinfo{year}{1960}).

\bibitem[{\citenamefont{Morrissey et~al.}(1994)\citenamefont{Morrissey,
  Benenson, and Friedman}}]{morrissey1}
\bibinfo{author}{\bibfnamefont{D.~J.} \bibnamefont{Morrissey}},
  \bibinfo{author}{\bibfnamefont{W.}~\bibnamefont{Benenson}}, \bibnamefont{and}
  \bibinfo{author}{\bibfnamefont{W.}~\bibnamefont{Friedman}},
  \bibinfo{journal}{Annual Review of Nuclear and Particle Science}
  \textbf{\bibinfo{volume}{44}}, \bibinfo{pages}{27} (\bibinfo{year}{1994}).

\bibitem[{\citenamefont{Gulminelli}(2004)}]{gulminelli1}
\bibinfo{author}{\bibfnamefont{F.}~\bibnamefont{Gulminelli}},
  \bibinfo{journal}{Ann. Phys. Fr} \textbf{\bibinfo{volume}{29}},
  \bibinfo{pages}{1} (\bibinfo{year}{2004}).

\bibitem[{\citenamefont{Gross}(2004)}]{gross1}
\bibinfo{author}{\bibfnamefont{D.~H.~E.} \bibnamefont{Gross}},
  \bibinfo{journal}{Entropy} \textbf{\bibinfo{volume}{6}}, \bibinfo{pages}{158}
  (\bibinfo{year}{2004}).

\bibitem[{\citenamefont{Gross}(1997)}]{gross2}
\bibinfo{author}{\bibfnamefont{D.}~\bibnamefont{Gross}},
  \bibinfo{journal}{Physics Reports} \textbf{\bibinfo{volume}{279}},
  \bibinfo{pages}{119 } (\bibinfo{year}{1997}).

\bibitem[{\citenamefont{Borderie and Rivet}(2008)}]{borderie1}
\bibinfo{author}{\bibfnamefont{B.}~\bibnamefont{Borderie}} \bibnamefont{and}
  \bibinfo{author}{\bibfnamefont{M.}~\bibnamefont{Rivet}},
  \bibinfo{journal}{Progress in Particle and Nuclear Physics}
  \textbf{\bibinfo{volume}{61}}, \bibinfo{pages}{551 } (\bibinfo{year}{2008}).

\bibitem[{\citenamefont{Kelic et~al.}(2006)\citenamefont{Kelic, Natowitz, and
  Schmidt}}]{kelic1}
\bibinfo{author}{\bibfnamefont{A.}~\bibnamefont{Kelic}},
  \bibinfo{author}{\bibfnamefont{J.~B.} \bibnamefont{Natowitz}},
  \bibnamefont{and} \bibinfo{author}{\bibfnamefont{K.~H.}
  \bibnamefont{Schmidt}}, \bibinfo{journal}{Eur. Phys. J.}
  \textbf{\bibinfo{volume}{A30}}, \bibinfo{pages}{203} (\bibinfo{year}{2006}).

\bibitem[{\citenamefont{Toepffer and Wong}(1982)}]{toepffer1}
\bibinfo{author}{\bibfnamefont{C.}~\bibnamefont{Toepffer}} \bibnamefont{and}
  \bibinfo{author}{\bibfnamefont{C.-Y.} \bibnamefont{Wong}},
  \bibinfo{journal}{Phys. Rev. C} \textbf{\bibinfo{volume}{25}},
  \bibinfo{pages}{1018} (\bibinfo{year}{1982}).

\bibitem[{\citenamefont{Cugnon}(1984)}]{cugnon1}
\bibinfo{author}{\bibfnamefont{J.}~\bibnamefont{Cugnon}},
  \bibinfo{journal}{Physics Letters B} \textbf{\bibinfo{volume}{135}},
  \bibinfo{pages}{374 } (\bibinfo{year}{1984}).

\bibitem[{\citenamefont{Jouan et~al.}(1991)}]{jouan1}
\bibinfo{author}{\bibfnamefont{D.}~\bibnamefont{Jouan}} \bibnamefont{et~al.},
  \bibinfo{journal}{Zeitschrift f{\"u}r Physik A Hadrons and Nuclei}
  \textbf{\bibinfo{volume}{340}}, \bibinfo{pages}{63} (\bibinfo{year}{1991}).

\bibitem[{\citenamefont{Borderie et~al.}(1999)}]{borderie2}
\bibinfo{author}{\bibfnamefont{B.}~\bibnamefont{Borderie}}
  \bibnamefont{et~al.}, \bibinfo{journal}{The European Physical Journal A -
  Hadrons and Nuclei} \textbf{\bibinfo{volume}{6}}, \bibinfo{pages}{197}
  (\bibinfo{year}{1999}).

\bibitem[{\citenamefont{{Bonche} et~al.}(1984)\citenamefont{{Bonche}, {Levit},
  and {Vautherin}}}]{bonche1}
\bibinfo{author}{\bibfnamefont{P.}~\bibnamefont{{Bonche}}},
  \bibinfo{author}{\bibfnamefont{S.}~\bibnamefont{{Levit}}}, \bibnamefont{and}
  \bibinfo{author}{\bibfnamefont{D.}~\bibnamefont{{Vautherin}}},
  \bibinfo{journal}{Nucl. Phys. A} \textbf{\bibinfo{volume}{427}},
  \bibinfo{pages}{278} (\bibinfo{year}{1984}).

\bibitem[{\citenamefont{{Levit} and {Bonche}}(1985)}]{levit1}
\bibinfo{author}{\bibfnamefont{S.}~\bibnamefont{{Levit}}} \bibnamefont{and}
  \bibinfo{author}{\bibfnamefont{P.}~\bibnamefont{{Bonche}}},
  \bibinfo{journal}{Nucl. Phys. A} \textbf{\bibinfo{volume}{437}},
  \bibinfo{pages}{426} (\bibinfo{year}{1985}).

\bibitem[{\citenamefont{{Besprosvany} and {Levit}}(1989)}]{besprosvany1}
\bibinfo{author}{\bibfnamefont{J.}~\bibnamefont{{Besprosvany}}}
  \bibnamefont{and} \bibinfo{author}{\bibfnamefont{S.}~\bibnamefont{{Levit}}},
  \bibinfo{journal}{Phys. Lett. B} \textbf{\bibinfo{volume}{217}},
  \bibinfo{pages}{1} (\bibinfo{year}{1989}).

\bibitem[{\citenamefont{{Pochodzalla} et~al.}(1995)}]{pochodzalla1}
\bibinfo{author}{\bibfnamefont{J.}~\bibnamefont{{Pochodzalla}}}
  \bibnamefont{et~al.}, \bibinfo{journal}{Phys. Rev. Lett.}
  \textbf{\bibinfo{volume}{75}}, \bibinfo{pages}{1040} (\bibinfo{year}{1995}).

\bibitem[{\citenamefont{Natowitz et~al.}(2002)}]{natowitz1}
\bibinfo{author}{\bibfnamefont{J.~B.} \bibnamefont{Natowitz}}
  \bibnamefont{et~al.}, \bibinfo{journal}{Phys. Rev.}
  \textbf{\bibinfo{volume}{C65}}, \bibinfo{pages}{034618}
  (\bibinfo{year}{2002}).

\bibitem[{\citenamefont{{Le F{\`e}vre} et~al.}(1999)\citenamefont{{Le
  F{\`e}vre}, {Schapiro}, and {Chbihi}}}]{lefevre1}
\bibinfo{author}{\bibfnamefont{A.}~\bibnamefont{{Le F{\`e}vre}}},
  \bibinfo{author}{\bibfnamefont{O.}~\bibnamefont{{Schapiro}}},
  \bibnamefont{and} \bibinfo{author}{\bibfnamefont{A.}~\bibnamefont{{Chbihi}}},
  \bibinfo{journal}{Nucl. Phys. A} \textbf{\bibinfo{volume}{657}},
  \bibinfo{pages}{446} (\bibinfo{year}{1999}).

\bibitem[{\citenamefont{Siwek et~al.}(1998)\citenamefont{Siwek, Durand,
  Gulminelli, and P\'eter}}]{siwek1}
\bibinfo{author}{\bibfnamefont{A.}~\bibnamefont{Siwek}},
  \bibinfo{author}{\bibfnamefont{D.}~\bibnamefont{Durand}},
  \bibinfo{author}{\bibfnamefont{F.}~\bibnamefont{Gulminelli}},
  \bibnamefont{and} \bibinfo{author}{\bibfnamefont{J.}~\bibnamefont{P\'eter}},
  \bibinfo{journal}{Phys. Rev. C} \textbf{\bibinfo{volume}{57}},
  \bibinfo{pages}{2507} (\bibinfo{year}{1998}),
  \urlprefix\url{https://link.aps.org/doi/10.1103/PhysRevC.57.2507}.

\bibitem[{\citenamefont{Charity et~al.}()}]{charity1}
\bibinfo{author}{\bibfnamefont{R.~J.} \bibnamefont{Charity}}
  \bibnamefont{et~al.}, \bibinfo{note}{computer Code GEMINI(unpublished)}.

\bibitem[{\citenamefont{Charity et~al.}(1988)}]{charity2}
\bibinfo{author}{\bibfnamefont{R.~J.} \bibnamefont{Charity}}
  \bibnamefont{et~al.}, \bibinfo{journal}{Nucl. Phys. A}
  \textbf{\bibinfo{volume}{483}}, \bibinfo{pages}{371} (\bibinfo{year}{1988}).

\bibitem[{\citenamefont{Nicolis et~al.}(1992)}]{nicolis1}
\bibinfo{author}{\bibfnamefont{N.~G.} \bibnamefont{Nicolis}}
  \bibnamefont{et~al.}, \bibinfo{journal}{Phys. Rev. C}
  \textbf{\bibinfo{volume}{45}}, \bibinfo{pages}{2393} (\bibinfo{year}{1992}).

\bibitem[{\citenamefont{Durand}(1992)}]{durand1}
\bibinfo{author}{\bibfnamefont{D.}~\bibnamefont{Durand}},
  \bibinfo{journal}{Nuclear Physics A} \textbf{\bibinfo{volume}{541}},
  \bibinfo{pages}{266} (\bibinfo{year}{1992}).

\bibitem[{\citenamefont{Gruyer et~al.}(2015)}]{gruyer1}
\bibinfo{author}{\bibfnamefont{D.}~\bibnamefont{Gruyer}} \bibnamefont{et~al.}
  (\bibinfo{collaboration}{INDRA Collaboration}), \bibinfo{journal}{Phys. Rev.
  C} \textbf{\bibinfo{volume}{92}}, \bibinfo{pages}{064606}
  (\bibinfo{year}{2015}).

\bibitem[{\citenamefont{Vient et~al.}(2002)}]{Vient1}
\bibinfo{author}{\bibfnamefont{E.}~\bibnamefont{Vient}} \bibnamefont{et~al.},
  \bibinfo{journal}{Nuclear Physics A} \textbf{\bibinfo{volume}{700}},
  \bibinfo{pages}{555 } (\bibinfo{year}{2002}).

\bibitem[{\citenamefont{Copinet}(1990)}]{copinet1}
\bibinfo{author}{\bibfnamefont{N.}~\bibnamefont{Copinet}}, Ph.D. thesis,
  \bibinfo{school}{Universit\'e de Caen} (\bibinfo{year}{1990}),
  \bibinfo{note}{rX-1328, GANIL T 90}.

\bibitem[{\citenamefont{Steckmeyer et~al.}(2001)}]{steck1}
\bibinfo{author}{\bibfnamefont{J.~C.} \bibnamefont{Steckmeyer}}
  \bibnamefont{et~al.}, \bibinfo{journal}{Nuclear Physics A}
  \textbf{\bibinfo{volume}{686}}, \bibinfo{pages}{537 } (\bibinfo{year}{2001}).

\bibitem[{\citenamefont{Pouthas et~al.}(1995)}]{Pouthas1}
\bibinfo{author}{\bibfnamefont{J.}~\bibnamefont{Pouthas}} \bibnamefont{et~al.},
  \bibinfo{journal}{Nuclear Instruments and Methods in Physics Research Section
  A: Accelerators, Spectrometers, Detectors and Associated Equipment}
  \textbf{\bibinfo{volume}{357}}, \bibinfo{pages}{418 } (\bibinfo{year}{1995}).

\bibitem[{\citenamefont{Vient}(2006)}]{Vient2}
\bibinfo{author}{\bibfnamefont{E.}~\bibnamefont{Vient}},
  \bibinfo{type}{M\'emoire d'habilitation \`a diriger des recherches},
  \bibinfo{school}{Universit\'e de Caen} (\bibinfo{year}{2006}).

\bibitem[{\citenamefont{Parker et~al.}(1991)}]{parker1}
\bibinfo{author}{\bibfnamefont{W.~E.} \bibnamefont{Parker}}
  \bibnamefont{et~al.}, \bibinfo{journal}{Phys. Rev. C}
  \textbf{\bibinfo{volume}{44}}, \bibinfo{pages}{774} (\bibinfo{year}{1991}).

\bibitem[{\citenamefont{Louis C.~Vaz and Alexander}(1984)}]{vaz1}
\bibinfo{author}{\bibfnamefont{L.}~\bibnamefont{Louis C.~Vaz}}
  \bibnamefont{and}
  \bibinfo{author}{\bibfnamefont{J.}~\bibnamefont{Alexander}},
  \bibinfo{journal}{Zeitschrift f\"{u}r Physik A Hadrons and Nuclei}
  \textbf{\bibinfo{volume}{318}}, \bibinfo{pages}{231} (\bibinfo{year}{1984}).

\bibitem[{\citenamefont{Ajitanand et~al.}(1986)\citenamefont{Ajitanand, Lacey,
  Peaslee, Duek, and Alexander}}]{ajitanand1}
\bibinfo{author}{\bibfnamefont{N.}~\bibnamefont{Ajitanand}},
  \bibinfo{author}{\bibfnamefont{R.}~\bibnamefont{Lacey}},
  \bibinfo{author}{\bibfnamefont{G.}~\bibnamefont{Peaslee}},
  \bibinfo{author}{\bibfnamefont{E.}~\bibnamefont{Duek}}, \bibnamefont{and}
  \bibinfo{author}{\bibfnamefont{J.~M.} \bibnamefont{Alexander}},
  \bibinfo{journal}{Nuclear Instruments and Methods in Physics Research Section
  A: Accelerators, Spectrometers, Detectors and Associated Equipment}
  \textbf{\bibinfo{volume}{243}}, \bibinfo{pages}{111 } (\bibinfo{year}{1986}).

\bibitem[{\citenamefont{Charity et~al.}(2001)}]{charity3}
\bibinfo{author}{\bibfnamefont{R.~J.} \bibnamefont{Charity}}
  \bibnamefont{et~al.}, \bibinfo{journal}{Phys. Rev. C}
  \textbf{\bibinfo{volume}{63}}, \bibinfo{pages}{024611}
  (\bibinfo{year}{2001}).

\bibitem[{\citenamefont{Wada et~al.}(1989)}]{wada1}
\bibinfo{author}{\bibfnamefont{R.}~\bibnamefont{Wada}} \bibnamefont{et~al.},
  \bibinfo{journal}{Phys. Rev. C} \textbf{\bibinfo{volume}{39}},
  \bibinfo{pages}{497} (\bibinfo{year}{1989}).

\bibitem[{\citenamefont{Hagel et~al.}(1988)}]{hagel1}
\bibinfo{author}{\bibfnamefont{K.}~\bibnamefont{Hagel}} \bibnamefont{et~al.},
  \bibinfo{journal}{Nuclear Physics A} \textbf{\bibinfo{volume}{486}},
  \bibinfo{pages}{429 } (\bibinfo{year}{1988}).

\bibitem[{\citenamefont{{W.H. Press, S.A. Teukolsky, W.T. Vetterling, B.P.
  Flannery}}(1992)}]{hpress1}
\bibinfo{author}{\bibnamefont{{W.H. Press, S.A. Teukolsky, W.T. Vetterling,
  B.P. Flannery}}}, \emph{\bibinfo{title}{Numerical Recipes in Fortran 77,
  \\the Art of scientific computing}} (\bibinfo{publisher}{Cambridge University
  Press}, \bibinfo{year}{1992}), vol.~\bibinfo{volume}{1}, p.
  \bibinfo{pages}{614}, \bibinfo{edition}{2nd} ed.

\bibitem[{\citenamefont{Hauser and Feshbach}(1952)}]{hauser1}
\bibinfo{author}{\bibfnamefont{W.}~\bibnamefont{Hauser}} \bibnamefont{and}
  \bibinfo{author}{\bibfnamefont{H.}~\bibnamefont{Feshbach}},
  \bibinfo{journal}{Phys. Rev.} \textbf{\bibinfo{volume}{87}},
  \bibinfo{pages}{366} (\bibinfo{year}{1952}).

\bibitem[{\citenamefont{Alexander et~al.}(1990)\citenamefont{Alexander, Magda,
  and Landowne}}]{alexander1}
\bibinfo{author}{\bibfnamefont{J.~M.} \bibnamefont{Alexander}},
  \bibinfo{author}{\bibfnamefont{M.~T.} \bibnamefont{Magda}}, \bibnamefont{and}
  \bibinfo{author}{\bibfnamefont{S.}~\bibnamefont{Landowne}},
  \bibinfo{journal}{Phys. Rev. C} \textbf{\bibinfo{volume}{42}},
  \bibinfo{pages}{1092} (\bibinfo{year}{1990}).

\bibitem[{\citenamefont{Nguyen}(1999)}]{nguyen1}
\bibinfo{author}{\bibfnamefont{A.}~\bibnamefont{Nguyen}}, Ph.D. thesis,
  \bibinfo{school}{Universit\'e de Caen} (\bibinfo{year}{1999}).

\bibitem[{\citenamefont{Bohr and Wheeler}(1939)}]{bohr2}
\bibinfo{author}{\bibfnamefont{N.}~\bibnamefont{Bohr}} \bibnamefont{and}
  \bibinfo{author}{\bibfnamefont{J.~A.} \bibnamefont{Wheeler}},
  \bibinfo{journal}{Phys. Rev.} \textbf{\bibinfo{volume}{56}},
  \bibinfo{pages}{426} (\bibinfo{year}{1939}).

\bibitem[{\citenamefont{Kramers}(1940)}]{kramers1}
\bibinfo{author}{\bibfnamefont{H.~A.} \bibnamefont{Kramers}},
  \bibinfo{journal}{Physica} \textbf{\bibinfo{volume}{7}}
  (\bibinfo{year}{1940}).

\end{thebibliography}

\end{document}